\documentclass[acmlarge,screen]{acmart}

\AtBeginDocument{%
  \providecommand\BibTeX{{%
    \normalfont B\kern-0.5em{\scshape i\kern-0.25em b}\kern-0.8em\TeX}}}

\setcopyright{acmcopyright}
\copyrightyear{2022}
\acmYear{2022}
\acmDOI{}

\acmConference[]{}{}{}
\acmPrice{15.00}
\acmISBN{}

\usepackage{caption}
\usepackage{subcaption}
\usepackage{tabularx}
\usepackage{algorithm}
\usepackage{algpseudocode}
\usepackage{xspace}

\usepackage[utf8]{inputenc}
\usepackage{ulem}
\usepackage{threeparttable}

\newcommand*{\systemname}{VReHab\xspace}

\begin{document}

%
\title[Facilitating Self-monitored Physical Rehabilitation with Virtual Reality and Haptic feedback]{Facilitating Self-monitored Physical Rehabilitation with Virtual Reality and Haptic feedback}

\author{Yu Jiang}
\authornote{Both authors contributed equally to this research.}
\email{jiangyu0927@gmail.com}
\orcid{1234-5678-9012}
\author{Zhipeng Li}
\authornotemark[1]
\email{}
\affiliation{%
  \institution{Department of Science and Technology, Tsinghua University}
  \streetaddress{P.O. Box 1212}
  \city{Beijing}
  \state{Ohio}
  \country{China}
  \postcode{43017-6221}
}

\author{Ziyue Dang}
\affiliation{%
  \institution{Department of Science and Technology, Tsinghua University}
  \streetaddress{1 Th{\o}rv{\"a}ld Circle}
  \city{Hekla}
  \country{China}}
\email{larst@affiliation.org}

\author{Yuntao Wang}
\affiliation{%
  \institution{Department of Science and Technology, Tsinghua University}
  \city{Rocquencourt}
  \country{China}
}

\author{Yukang Yan}
\affiliation{%
 \institution{Department of Science and Technology, Tsinghua University}
 \streetaddress{Rono-Hills}
 \city{Doimukh}
 \state{Arunachal Pradesh}
 \country{China}}

\author{Y ZHANG}
\affiliation{%
  \institution{Peking University Third Hospital}
  \streetaddress{30 Shuangqing Rd}
  \city{Haidian Qu}
  \state{Beijing Shi}
  \country{China}}

\author{Xinguang Wang}
\affiliation{%
  \institution{Department of Orthopaedics, Peking University Third Hospital}
  \streetaddress{8600 Datapoint Drive}
  \city{San Antonio}
  \state{Texas}
  \country{China}
  \postcode{78229}}
\email{cpalmer@prl.com}

\author{Yansong Li}
\affiliation{%
  \institution{Health Science Center, Peking University}
  \streetaddress{1 Th{\o}rv{\"a}ld Circle}
  \city{Hekla}
  \country{China}}
\email{jsmith@affiliation.org}

\author{Mouwang Zhou}
\affiliation{%
  \institution{Peking University Third Hospital}
  \city{New York}
  \country{China}}
\email{jpkumquat@consortium.net}

\author{Hua Tian}
\affiliation{%
  \institution{Department of Orthopaedics, Peking University Third Hospital}
  \city{New York}
  \country{China}}
\email{jpkumquat@consortium.net}

\author{Yuanchun Shi}
\affiliation{%
  \institution{Department of Computer science and Technology, Tsinghua University}
  \city{New York}
  \country{China}}
\email{jpkumquat@consortium.net}

\renewcommand{\shortauthors}{Jiang and Li, et al.}

\begin{abstract}
Physical rehabilitation is essential to recovery from joint replacement operations. As a representation, total knee arthroplasty (TKA) requires patients to conduct intensive physical exercises to regain the knee's range of motion and muscle strength.
However, current joint replacement physical rehabilitation methods rely highly on therapists for supervision, and existing computer-assisted systems lack consideration for enabling self-monitoring, making at-home physical rehabilitation difficult.
In this paper, we investigated design recommendations that would enable self-monitored rehabilitation through clinical observations and focus group interviews with doctors and therapists.
With this knowledge, we further explored Virtual Reality(VR)-based visual presentation and supplemental haptic motion guidance features in our implementation \systemname, a self-monitored and multimodal physical rehabilitation system with VR and vibrotactile and pneumatic feedback in a TKA rehabilitation context.
We found that the third point of view real-time reconstructed motion on a virtual avatar overlaid with the target pose effectively provides motion awareness and guidance while haptic feedback helps enhance users' motion accuracy and stability.
Finally, we implemented \systemname to facilitate self-monitored post-operative exercises and validated its effectiveness through a clinical study with 10 patients.

\end{abstract}



\begin{teaserfigure}
    \centering
  \includegraphics[width=0.9\textwidth]{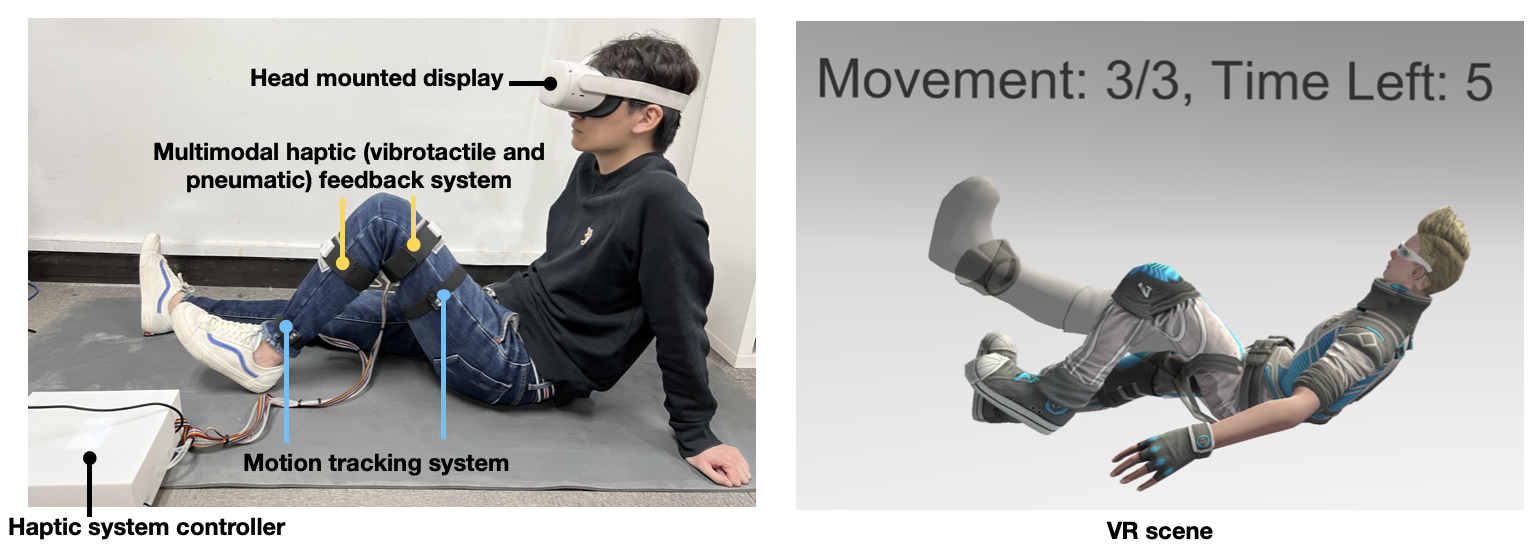}
  \caption{VReHab - our implementation of a multimodal and and self-monitored physical rehabilitation system with Virtual Reality and haptic feedback for postoperative TKA patients based on the proposed design guidelines.}
  \label{fig:teaser}
\end{teaserfigure}

\maketitle

\section{Introduction}
Long-term physical rehabilitation is essential for the recovery of motor skills and health-related quality of life after injuries or operations~\cite{jung_20}. 
Rehabilitation post total knee arthroplasty (TKA) operation focuses on the range of motion and muscle strength exercises which requires patients to be under close supervision.
Despite TKA's rising case number in China, current at-home postoperative physical rehabilitation fails to effectively aid recovery due to the lack of therapeutically appropriate guidance and supervision~\cite{feng_20}.
Past studies proposed computer-assisted systems that enabled at-home physical rehabilitation.
Some leveraged games to motivate engagement in exercises~\cite{DBLP:journals/imwut/UzorB17, DBLP:journals/tochi/JungPMPRKL20, del_12, dunne_10, luo_10, halic_14, tsekleves_14, villiger_11, jung_20}, or focused on journaling users' motions with unobtrusive on-body sensors~\cite{spina_13, ferreira_14, ongvisa_15, tsekleves_14, fung_12, timmermans_10, steffen_11, fortino_14}. 
Yet these works lack consideration in enabling patients to monitor their exercises which is crucial in TKA rehabilitation.
Works that reconstructed users' motion~\cite{ayoade_14, yeh_12} to enhance self-perception provided limited guidance and were restricted to visual presentation with traditional screen displays, which are hard to mobilize when performing large-scale motions.
We thus set out to propose a multimodal, self-monitored, and computer-assisted physical rehabilitation system in which patients take on the role of supervising their movements and rely on the computer for accurate and effective motion guidance.

We conducted focus group surveys with doctors and therapists and observations of clinical TKA rehabilitation sessions to thoroughly understand the needs and challenges of self-monitored TKA physical rehabilitation.
We found it's crucial that such a system provides motion awareness and motion guidance based on accurate tracking, involves therapists asynchronously, provides multimodal supervision, achieves automated progress tracking, and has mobility. 
Based on these design recommendations, we leveraged VR with head-mounted displays (HMDs) for mobility and immersion in visual displays and haptic feedback for supplemental and accurate motion guidance to implement a self-monitored physical rehabilitation system \systemname based on VR and haptic feedback.
Then through a series of user studies in a lab setting, we further characterized the visual and haptic feedback in \systemname by evaluating different visual and haptic factors' influence on the time needed to understand instructions, angle accuracy, and motion stability.
Specifically, we evaluated points of view and the presence of a virtual supporting hand on the leg for the visual feedback; and we evaluated vibrotactile and pneumatic feedback for the haptic feedback. 
The results showed that the third point of view in VR without the virtual hand provides better motion awareness and guidance, and haptic feedback enhances motion accuracy and stability.
We found vibrotactile feedback intense and instantaneous yet numbing while pneumatic feedback soothing and pseudo-kinesthetic but slow in actuation.
Based on the findings, we implemented \systemname with VR and vibrotactile and pneumatic feedback for self-monitored TKA rehabilitation.
We investigated the effectiveness of \systemname in facilitating self-monitored TKA physical rehabilitation through a clinical evaluation (N = 10) and semi-structured interviews. 
The result revealed that our system could significantly improve motion stability while incurring a slight increase in instruction understanding time. 
The semi-structured interviews showed that users were willing and confident to use our system for self-monitored physical rehabilitation at home.

This paper's main contributions are three-folded:
\begin{itemize}
    \item We investigated crucial elements and proposed design recommendations of a self-monitored, computer-assisted physical rehabilitation system through clinical session observations and focus group interviews.
    \item We designed and characterized a self-monitored and multimodal rehabilitation system, \systemname, through evaluating factors of the VR(HMD) based visual (point of view and virtual authority) and haptic (vibrotactile and pneumatic) feedback and proposed a vibrotactile and pneumatic combined haptic feedback mechanism.
    \item We evaluated the effectiveness of \systemname and patients' willingness to use \systemname for self-monitored physical rehabilitation in a clinical evaluation (N = 10).
\end{itemize}

\section{Related Work}

\subsection{Self-monitored physical rehabilitation}
Self-monitored physical rehabilitation systems allow patients to monitor their performance during the practice by providing real-time motion awareness and guidance~\cite{piqueras_13, ayoade_14, yeh_12}.
Past studies leveraged smartphones~\cite{spina_13, ferreira_14, ongvisa_15}, Kinect~\cite{sadihov_13}, Wii controllers~\cite{tsekleves_14, fung_12}, gyroscopes embedded wearables~\cite{timmermans_10, steffen_11, fortino_14}, and force-sensitive motion tracking sensors~\cite{bin_12, halic_14} to track kinematic motions.
In providing real-time motion guidance based on the tracking, some studies presented only selected measurements, like the angle of a curved leg~\cite{piqueras_13, spina_13}. 
Live video~\cite{timmermans_10, bin_12, halic_14, luo_10} or audio~\cite{phyo_10} rehabilitation sessions at-home with a remote therapist have also been proposed. 
Physio@Home~\cite{tang_15} used multiple cameras to capture the patient's movements overlaid with pre-recorded videos.
Others provided 3d kinematic motions reconstructed from the sensors' data as visual feedback to ensure self-perception and awareness so patients can supervise themselves~\cite{ayoade_14, yeh_12}.

These proposed self-monitored rehabilitation systems either present limited information and guidance to users or require setting up extra equipment.
There also lacks an understanding of how to design a self-monitored rehabilitation system to incorporate the needs of the patients and the therapists.
We thus aim to fill this gap by investigating and proposing guidelines for designing self-monitored physical rehabilitation systems.

\vspace{-5mm}
\subsection{Unimodal motion guidance for physical rehabilitation}
\label{sec:unimodal_literature}
It's crucial to provide motion guidance for self-monitored physical rehabilitation.
Most of the previous studies focused on leveraging screen displays~\cite{timmermans_10, bin_12, halic_14, luo_10} to provide motion guidance.
With the improvement of immersive Virtual Reality (VR) technologies, more studies have explored using VR to facilitate physical rehabilitation and training~\cite{sveistrup_05, Grassini_20}.
\citeauthor{holden_02} developed a VR training system to enhance motor learning through augmented feedback by a virtual teacher~\cite{holden_02}.
\citeauthor{kim_99} developed a virtual cycling system to improve users' postural balance control~\cite{kim_99}. 
These research demonstrated that Head-Mounted Displays (HMDs) outperform desktop monitors in mobility, making them suitable for VR-based rehabilitation systems that involve large-scale movements. 

Some studies provided motion guidance based solely on haptic feedback including the use of vibrotactile~\cite{ding_13}, pneumatic~\cite{goto_18}, electrical muscle stimulation (EMS)~\cite{lopes2018adding, Lopes_17}, and skin stretch~\cite{chinello_18}.
\citeauthor{norman_14} guided the user's hand movements by providing skin stretch feedback on fingers~\cite{norman_14}.
\citeauthor{panchan_14} provided vibrotactile directional feedback to guide users towards target poses~\cite{panchan_14}. 

Other than visual and haptic feedback, several works investigating auditory feedback for physical rehabilitation~\cite{phyo_10, spina_13}.
However, providing unimodal feedback for motion guidance could be inefficient and imprecise due to the limitation of information presentation bandwidth.
Therefore we proposed to leverage multimodal feedback to provide motion guidance for self-monitored physical rehabilitation. 

\vspace{-5mm}
\subsection{Multimodal motion guidance for physical rehabilitation}
\label{sec:multimodal_literature}
In terms of providing multimodal feedback for motion guidance, it has been common to couple visual displays with haptic feedback devices.
Vibrotactile actuators have been leveraged together with visual displays to indicate the direction and extent to which the users need to move~\cite{del_12, kapur_10, Yang_02}. 
\citeauthor{steinisch_13}~\cite{steinisch_13} and \citeauthor{montagner_07}~\cite{montagner_07} used exoskeleton haptic devices and developed VR applications to train different movement patterns for neuro-motor rehabilitation of upper limbs in stroke survivors.
Desktop mechanical arms such as Phantom~\cite{broeren_04} and Novint Falcon~\cite{yeh_17, zhaohong_10} have been used to motivate motor training in stroke rehabilitation.
Others used gloves~\cite{jack_00, bortone_20, sadihov_13} to provide force feedback to fingers.
For lower-limb multimodal training, \citeauthor{riva_98} linked a gait-inducing exoskeleton to a HMD for post-spinal cord injuries physical rehabilitation~\cite{riva_98} and \citeauthor{fung_04} coupled a virtual environment with a self-paced treadmill for post-stroke locomotor training~\cite{fung_04}. 
\citeauthor{steffen_11}~\cite{steffen_11} and \citeauthor{rahman_15}~\cite{rahman_15} also explored using audio-visual instructions to guide users completing exercises.
The proposed exoskeleton devices are often immobile, costly, and difficult to set up, making them unfit for a self-monitored physical rehabilitation scenario. 

Furthermore, these proposed systems oftentimes lack characterization and evaluation, leaving their effectiveness to assist rehabilitation unknown. We thus aim to propose a lightweight, mobile, and low-cost multimodal self-monitored rehabilitation system based on the characterization of different modalities. We also aim to evaluate the effectiveness of our proposed system with a clinical study.

\vspace{-3mm}
\section{Understanding challenges and needs in self-monitored TKA rehabilitation}
\label{sec:interview}


\subsection{Current clinical and at-home TKA physical rehabilitation}
\label{observation}

\begin{figure}[htbp]
     \centering
     \begin{subfigure}{0.24\textwidth}
        \centering
        \includegraphics[width=\linewidth]{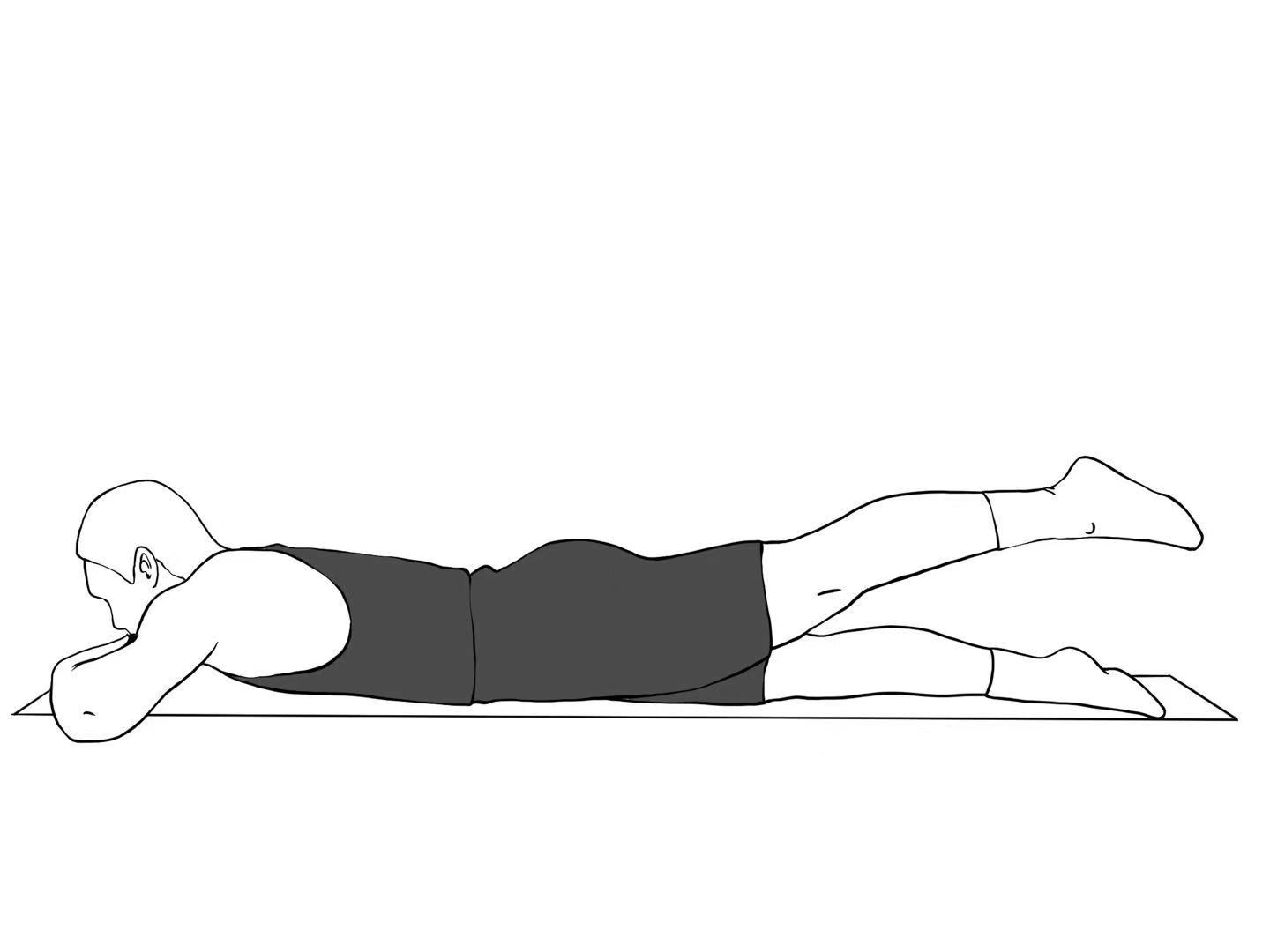}
        \caption{Straight leg raises (muscle strength exercise)}
        \label{fig:motionstraight}
     \end{subfigure}
     \hfill
     \begin{subfigure}{0.24\textwidth}
        \centering
        \includegraphics[width=\linewidth]{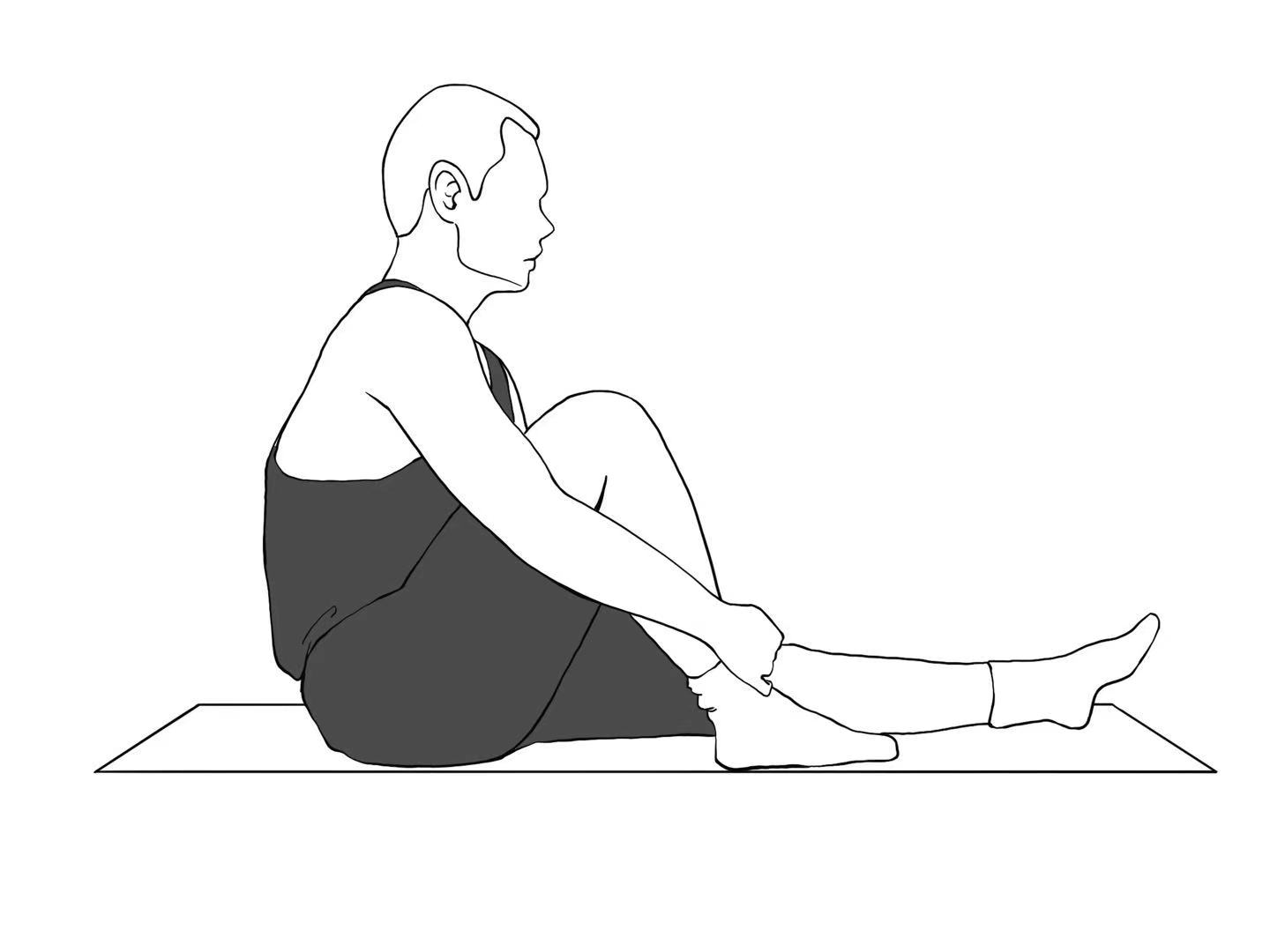}
        \caption{Bed-supported knee bends (range of motion exercise)}
        \label{fig:motionkneebend}
     \end{subfigure}
     \hfill
     \begin{subfigure}{0.24\textwidth}
        \centering
        \includegraphics[width=\linewidth]{figures/motion/prone_straight_leg_raises.jpeg}
        \caption{Prone straight leg raises (muscle strength exercise)}
        \label{fig:motionprone}
     \end{subfigure}
     \hfill
     \begin{subfigure}{0.24\textwidth}
        \centering
        \includegraphics[width=\linewidth]{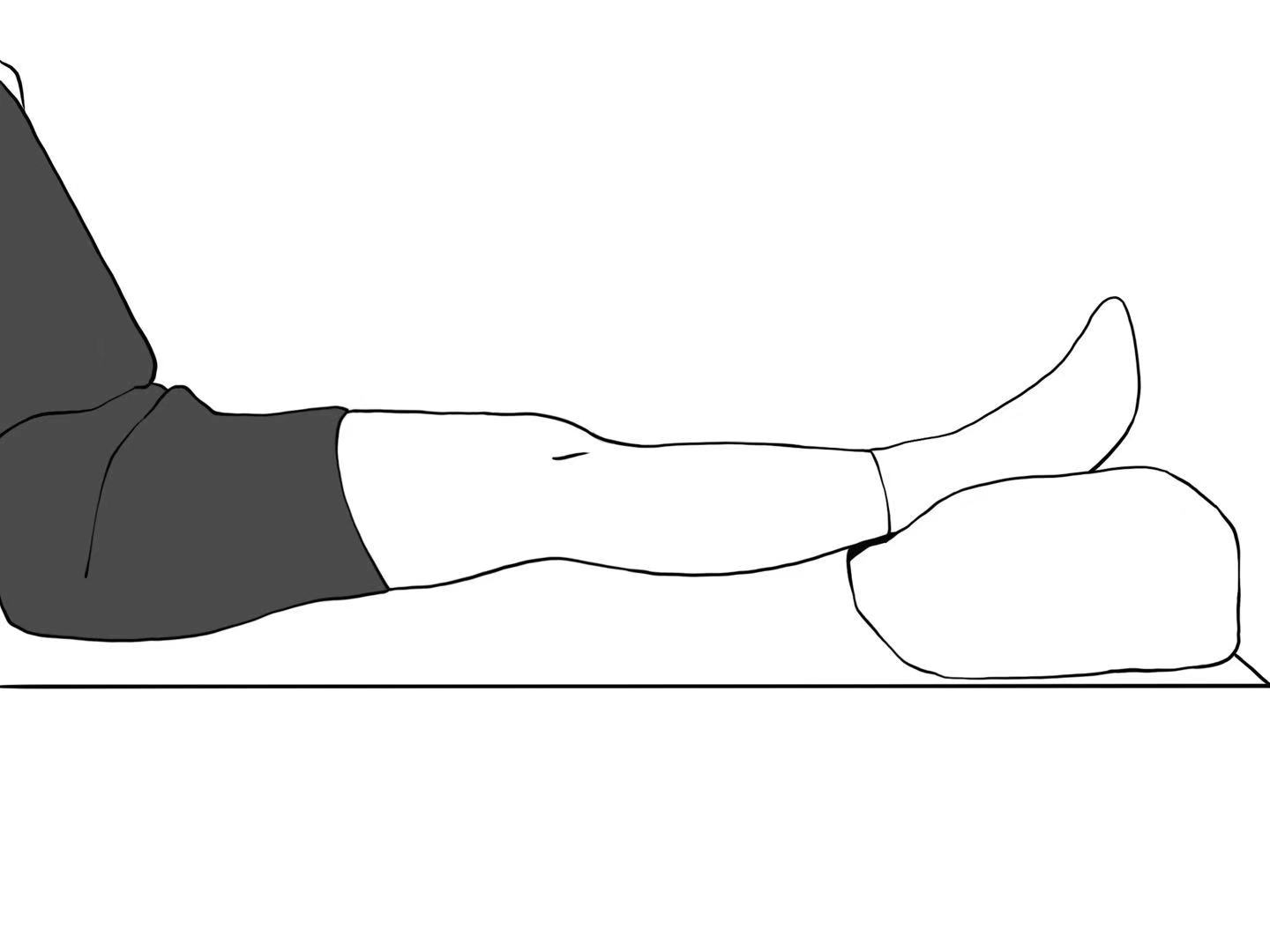}
        \caption{Knee extensions (range of motion exercise)}
        \label{fig:motionkneeextend}
     \end{subfigure}
     \caption{Four example poses in TKA physical rehabilitation that covers muscle strength and range of motion exercises. }
    \label{fig:poses}
\vspace{-5mm}
\end{figure}

To understand the challenges and needs in self-monitored TKA physical rehabilitation, we observed 2 patients (both third day after the surgery) in their daily clinical rehabilitation sessions at a local hospital. 
The therapist first recorded the injured legs' range of motion by measuring the knee extension and flexion angles with an angle gauge. 
Patients' self-reported level of pain was also recorded.
Based on this pre-session assessment, the therapist then prescribed the session's goals including the range of motion goals and the number of repetitions for each motion exercise.
Figure~\ref{fig:poses} illustrates the most important poses in a session, including \textbf{straight leg raises} (muscle strength exercise), \textbf{prone straight leg raises} (muscle strength exercise), \textbf{bed-supported knee bends} (range of motion exercise), and \textbf{knee extensions} (range of motion exercise).
The straight/prone leg raises require patients to raise their straightened leg until the angle from the calf/thigh to the bed is 30 degrees. 
The bed-supported knee bends require patients to sit upright on the bed then bend the injured leg towards the chest until the angle between the thigh and the calf is 90 degrees.
The knee extensions require that the angle between the thigh and the calf reaches 180 degrees.
Patients took 5-minute breaks between each pose exercise. 
The therapists vocally guided the patients through different poses. 

Therapist-guided physical rehabilitation at the hospital starts one day after the surgery and usually lasts 3 to 5 days with 2 sessions per day. 
After the patients are discharged from the hospital, long-term TKA rehabilitation, with the help of families or caregivers, is needed for effective recovery.
The same long-term and at-home TKA physical rehabilitation plan is prescribed one time to all discharged patients in form of a paper-based document. 
The document outlines patients' daily physical rehabilitation exercise requirements. 
The prescribed exercises are identical to those done at the hospital yet less intense to avoid re-hurting since the therapists' supervision is lacking.


\vspace{-3mm}
\subsection{Focus group surveys}
\label{sub:interview}
We then conducted two focus group surveys. Each group has one doctor from the orthopedics department (OD), one doctor from the physical rehabilitation department (RD), and one physical therapist (PT). 
During each focus group survey, the participants freely discussed what is needed for a therapeutically appropriate (i.e. professional and safe) and self-monitored TKA physical rehabilitation system.
Each group discussion lasted about 40 minutes and we collected 89 minutes of audio-recording in total.
We transcribed the recording and qualitatively coded their responses.
Three researchers worked separately for the open coding process and reached a consensus through discussions. 
The summarized design guidelines are listed and discussed below.

\vspace{-3mm}
\subsubsection{Providing motion awareness and guidance with accurate tracking}
\label{sec:motionguidance}
PT2 pointed out that accurate and real-time motion guidance is essential in postoperative self-monitored physical rehabilitation, which is currently lacking.
In a home setting without the therapist, patients are oftentimes \textit{"...unaware of their motion status or the target pose"} (RD1) and are frequently \textit{"...distracted if they see something more interesting"} (RD2)
This leads to ineffective exercises as patients may unintentionally fail to meet the requirement. 
OD1 suggested that \textit{"if the system can give visual or other forms of feedback to reflect patients' current motion, it would be much easier for self-monitoring."} 
Therefore, real-time movement feedback should be provided to users to enhance motion awareness.
Effective and concise motion guidance should also be provided to guide patients to achieve the target pose.

To achieve this, OD2 emphasized the importance of accurate and automated tracking of the patients' motion since the angle gauge used in pre-session assessments \textit{"has a 5-10 degree error and is time-consuming to use"} (OD1). PT1 and PT2 indicated that a tracking error within 5 degrees would be sufficient.

\vspace{-3mm}
\subsubsection{Keep therapists in the loop asynchronously}
\label{sec:keepintheloop}
All participants recognized that a self-monitored physical rehabilitation system would make remote or at-home session possible. 
It would also bring economics of scale as one therapist can now supervise many patients instead of having one-on-one sessions. 
PT1 further pointed out that \textit{"having summative reports with processed motion data from the sessions"} would help supervise multiple patients.
Based on the reports, therapists can timely adjust the prescribed physical rehabilitation plans. 
Knowing that their self-monitored physical rehabilitation sessions are remotely and asynchronously supervised by therapists can also be reassuring for patients who \textit{"...may feel that they are carrying out the sessions in a safe, professionally guided, and effective manner"}(PT2). 
This in turn enhances the patient's engagement with the physical rehabilitation.

\vspace{-3mm}
\subsubsection{Provide multimodal supervision}
\label{sec:authority}
The participants commented that patients' incompliance with the prescribed physical rehabilitation plan or "malingering" \cite{Burdea_03}, which describes patients purposefully do not exercise at their full capacity, is one of the crucial challenges of self-supervised physical rehabilitation. Malingering can be effectively reduced by supervision.

All participants agreed that in a clinical setting, an professional and supervisory figure (i.e. a therapist) can ensure the patients' adherence to a session. 
This is especially evident in postoperative TKA physical rehabilitation as the patients are mostly elders who are \textit{"...less independent and more reliant on the therapists"} (RD2). 
OD2 and RD1 indicated that the patients were more willing to adhere to clinical sessions because \textit{"they feel more obliged with a therapist by their side, knowing that the exercises they are doing are correct and safe."}. 
Thus a self-monitored physical rehabilitation system should consider the involvement of an authority character, e.g. a virtual therapist, who provides psychological supervision. 
We also observed that therapists leveraged haptic supervision by pushing or pulling the training legs if patients under- or over-performed.
PT1 added that this helped overcome patients' \textit{"fear of pain."} since patients usually don't exercise to their full potential because "they fear hurting themselves and tend to avoid the pain" (RD2).


\vspace{-3mm}
\subsubsection{Automated progress tracking throughout a session}
\label{sec:automated}
Our observation showed that patients went through several kinematically meaningful phases, specifically understanding, reaching, fine-tuning, holding, and retrieving, to complete a movement.
Postoperative TKA patients, who are usually elders, are often slow in understanding the vocal instructions due to their reduced cognitive capabilities. 
Therapists indicated that they \textit{"...hoped to deliver instructions through more effective and automated approaches to help expedite the understanding process"} (PT1, PT2).
The efficiency of a session thus can be improved by reducing \textbf{the time needed to understand instructions}. 
Upon understanding the instructions, patients then start to reach the target pose and fine-tune their poses if they over- or under-perform to ensure the quality and the safety of the training.
After the patients stabilize their poses, they need to hold the poses for 10 seconds.
Therapists indicated that they evaluated a patient's performance during a rehabilitation session subjectively in this holding phase by \textbf{how stable can the patient maintain a pose} and \textbf{the angular accuracy of the patient's pose compared to the target}.
After holding the poses, the patients can then retrieve their legs.

Currently, patients' movements are counted and timed by the therapists.
For at-home exercises, the doctors (RD1, RD2) commented that \textit{"it's hard and distracting for patients to conduct the exercises while counting and timing their movements at the same time."}
A self-monitored physical rehabilitation system thus should help alleviate the burden of progress tracking through automatically counting and timing, and offering flexibility in terms of the exercising pace.

\vspace{-3mm}
\subsubsection{Mobility of the entire system}
\label{sec:mobility}
The example poses demonstrated in Figure~\ref{fig:poses} require large-scale motions with the patients' head and body reorienting and repositioning for different poses.
Therapists (PT1, PT2) thus advised against immobile devices, such as fixed screen displays and heavy exoskeleton devices, as they would \textit{"...require others to help move the display/device to a new location for different poses."}
The mobility of the entire system is also crucial to facilitating self-monitored physical rehabilitation at unfixed locations.
\vspace{-3mm}
\section{\systemname System design and characterization} 
\label{sec:system_design}
This section presents the design and characterization \systemname for self-monitored and multimodal physical rehabilitation.

\vspace{-3mm}
\subsection{System Design}
\vspace{-1mm}
\subsubsection{Real-time Motion Tracking}
\label{sec:motiontracking}

We adopted two inertial measurement units (IMU) (HI221 from HiPNUC~\footnote{http://hipnuc.com/}). 
Each unit consists a MPU9250~\footnote{https://invensense.tdk.com/products/motion-tracking/9-axis/mpu-9250/} motion sensor, a micro-controller with Bluetooth for transmitting data wirelessly, and a built-in Kalman filter for accurate angle estimation. 
The IMUs work at 100 Hz frame rate. 
Each IMU sensor is embedded in a tight-fit 3d printed case adhered to a 2cm wide and 20cm long Velcro elastic tape (Fig~\ref{fig:imu}). 
We verified the IMU's robustness by attaching it to a robot arm programmed to move in six degrees of freedom for twenty minutes.
The absolute errors of the IMU's yaw, pitch, and roll angles were within 2 degrees, which were sufficient for \systemname.
The two IMUs are placed at the same lateral side of both the thigh and the ankle. 
We conducted a pilot study(N=3) in a lab setting with the IMU straps to verify the phases of a movement as observed in Section~\ref{sec:automated}. 
Fig~\ref{fig:segmentation} shows a color-coded example of a kinematically segmented straight leg raise movement with each phase exhibiting discernible patterns.

\vspace{-3mm}
\subsubsection{Automated progress tracking and segmentation}
\begin{figure}[htbp]
    \begin{minipage}{0.25\textwidth}
        \centering
        \includegraphics[width=\linewidth]{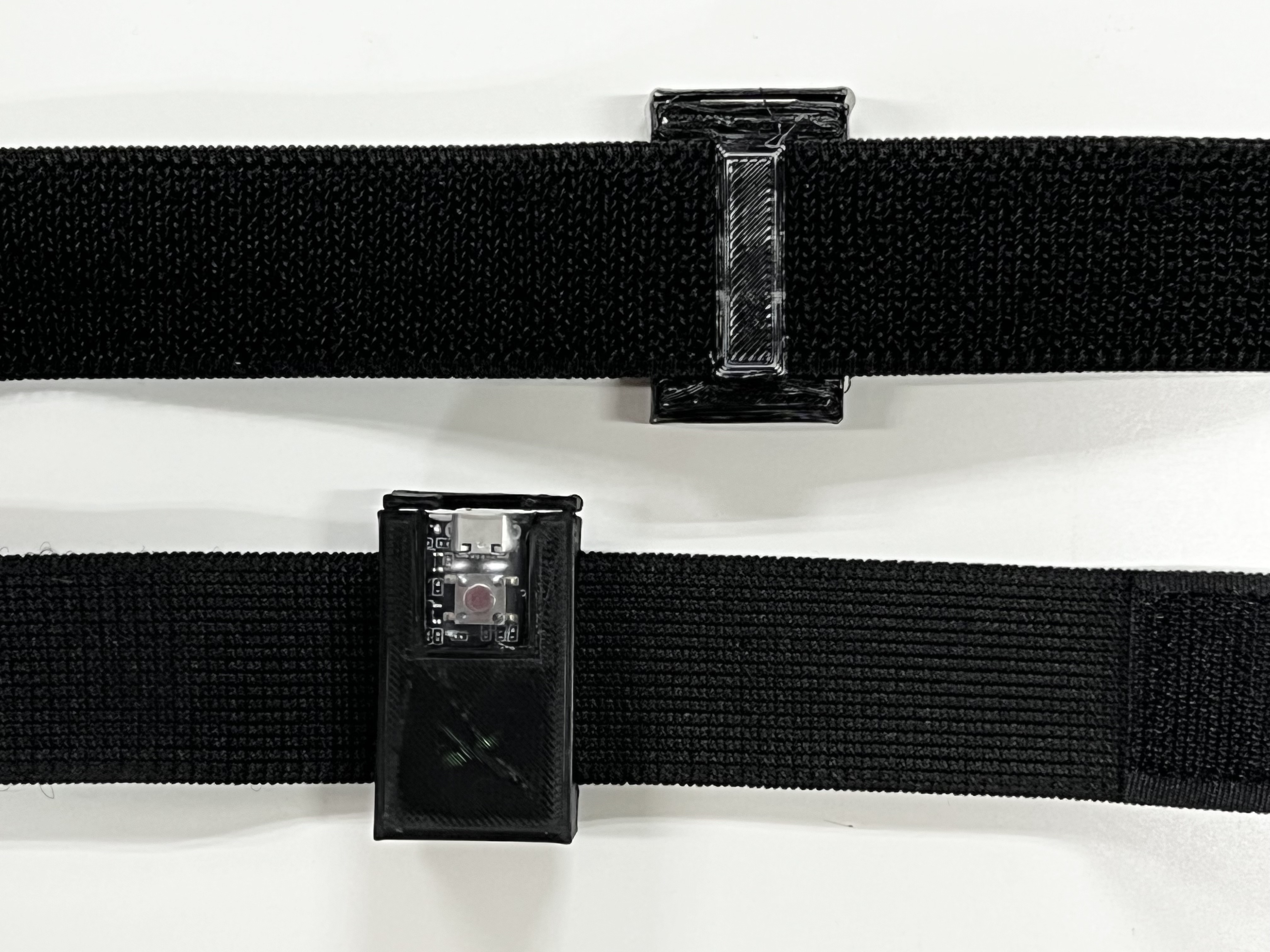}
        \caption{IMU sensors embedded in 3d printed cases and fastened on Velcro elastic straps. }
        \label{fig:imu}
    \end{minipage}
    \hspace{2em}
    \begin{minipage}{0.4\textwidth}
        \centering
        \includegraphics[width=\linewidth]{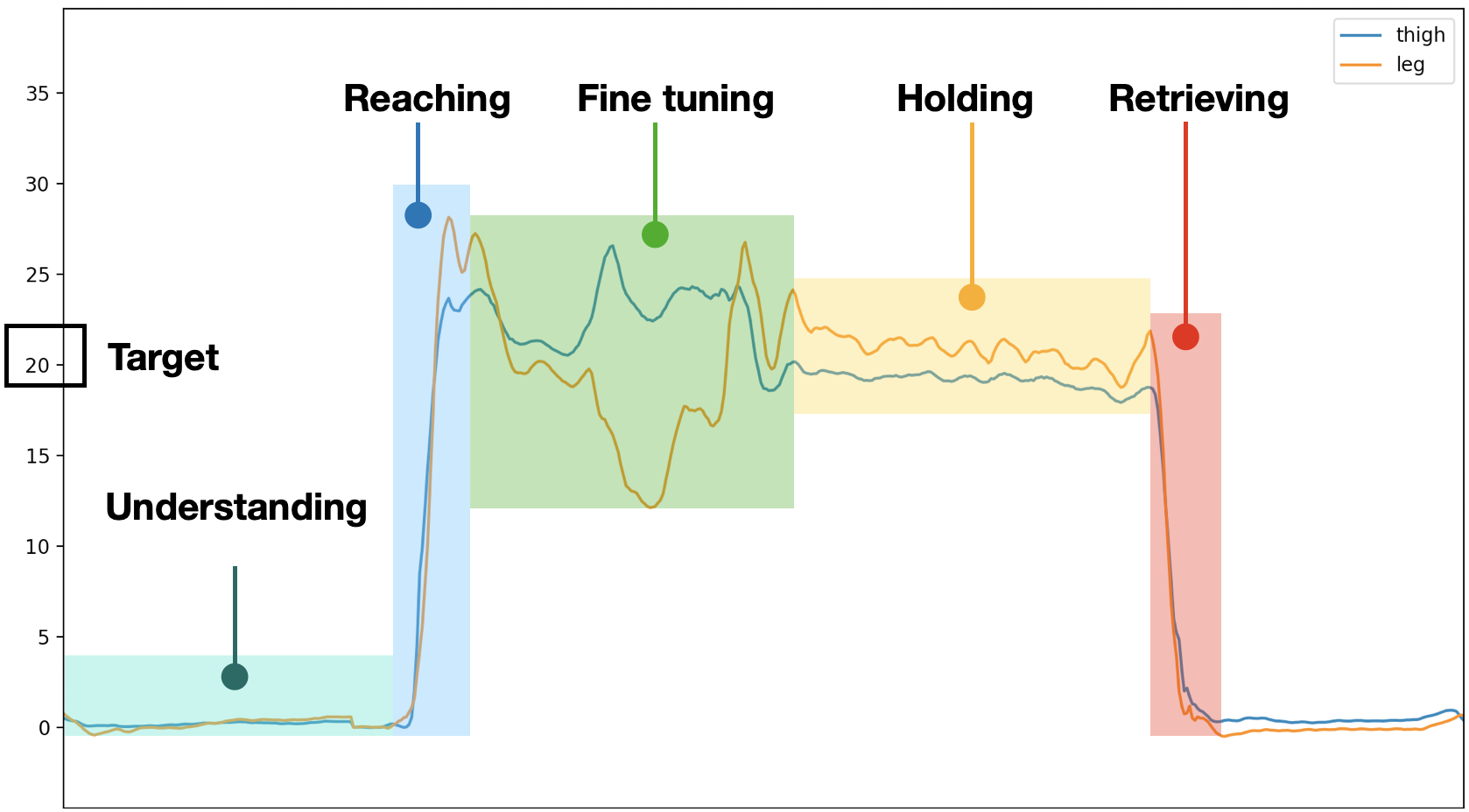}
        \caption{A movement was tracked and segmented into 5 kinematic phases.}
    \label{fig:segmentation}
    \end{minipage}
\vspace{-7mm}
\end{figure}

\systemname should track patients' movement through phases and should help time and count the motions (Sec~\ref{sec:automated}). 
This is achieved by using the observed angular changes in different phases (Fig~\ref{fig:segmentation}) to automatically determine the patients' current phase in real-time.
Based on this, \systemname can also set phase-specific goals and provide the according feedback.
Specifically, the system should provide intuitive and clear motion guidance during understanding.
Motion awareness and guidance should be provided during reaching to effectively guide patients. 
Supervision should be provided in the holding phase to help improve stability of a pose. 

\vspace{-3mm}
\subsubsection{Visual Feedback Design}
\label{sec:visual_design}


We adopted virtual reality (VR) through mobile head-mounted displays (HMDs) to provide visual feedback. 
An avatar in the VR represents the user and is displayed in a default white background. 
The reasons for adopting VR with HMDs in \systemname include: (1) mobility (Sec~\ref{sec:mobility}); (2) to avoid distractions by virtually re-constructing the user's environment; (3) to motivate users by providing immersion (Sec~\ref{sec:unimodal_literature}); (4) to provide motion awareness by reconstructing the user's movement on a virtual avatar  (Sec~\ref{sec:motionguidance}); and (5) to visualize the target pose on a target virtual avatar to provide motion guidance (Sec~\ref{sec:motionguidance}).
We projected the data obtained from the two IMU sensors onto the leg movement of an avatar in the virtual environment. 
For each exercise, a semi-transparent target avatar is overlaid onto the avatar to provide an intuitive visual representation of the target pose (Fig~\ref{fig:visual_conditions}).

\vspace{-3mm}
\subsubsection{Haptic Feedback Design}
\label{sec:hapticdesign}
Haptic feedback guides the user towards the target pose by reflecting the directional deviation of the current pose to the target pose.
It simulates a force that pushes or pulls the exercising leg to the target.
Haptic feedback is provided throughout the reaching phase to guide the patients to move in a certain direction; it's also provided in the fine-tuning and holding phase intermittently when the patients move out of the required range. 

The reasons for adopting haptic feedback in \systemname include: (1) to transform the visually subtle disparity between user and target avatars to obvious and accurate feedback to enhance motion awareness (Sec~\ref{sec:motionguidance}); (2) to instruct patients to move towards a certain direction (Sec~\ref{sec:motionguidance}); (3) to notify users that active action needs to be taken with external stimulus; and (4) to simulate a therapist pushing/pulling a patients' leg, which helps prevent malingering (Sec~\ref{sec:authority}).

\vspace{-3mm}
\subsection{Visual Feedback Characterization}
\label{sec:visual}
We then further characterize \systemname's visual feedback through investigating factors of interests.
Past studies showed that the third pov demonstrated the relative position of the user's legs to the surroundings~\cite{yu2020perspective}, while the first pov enhanced immersion and engagement in VR~\cite{fiorella2017s, boucheix2017reducing}.
We thus investigate how \textbf{point of view(pov)} affects visual information delivery in a physical rehabilitation scenario.
We also want to investigate the effect of the \textbf{presence of authority} on the performance and the mentality (i.e., stress) of the patients (Section~\ref{sec:authority}).

\vspace{-3mm}
\subsubsection{Design}
\begin{figure}[htbp]
     \centering
     \begin{subfigure}{0.24\textwidth}
        \centering
        \includegraphics[width=\linewidth]{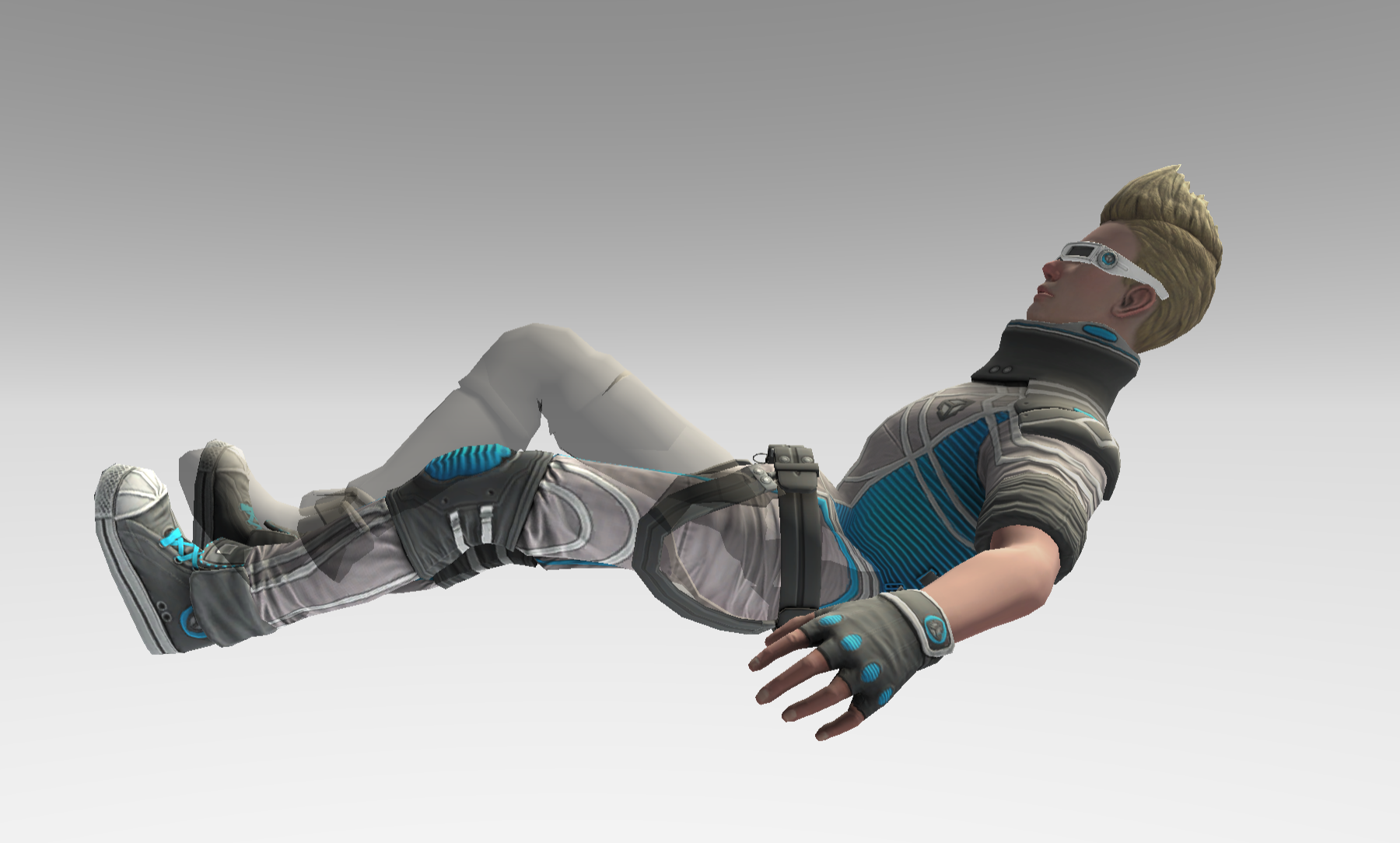}
        \caption{\textbf{TPH}: third point of view with hand clues}
        \label{fig:tph}
     \end{subfigure}
     \hfill
     \begin{subfigure}{0.24\textwidth}
        \centering
        \includegraphics[width=\linewidth]{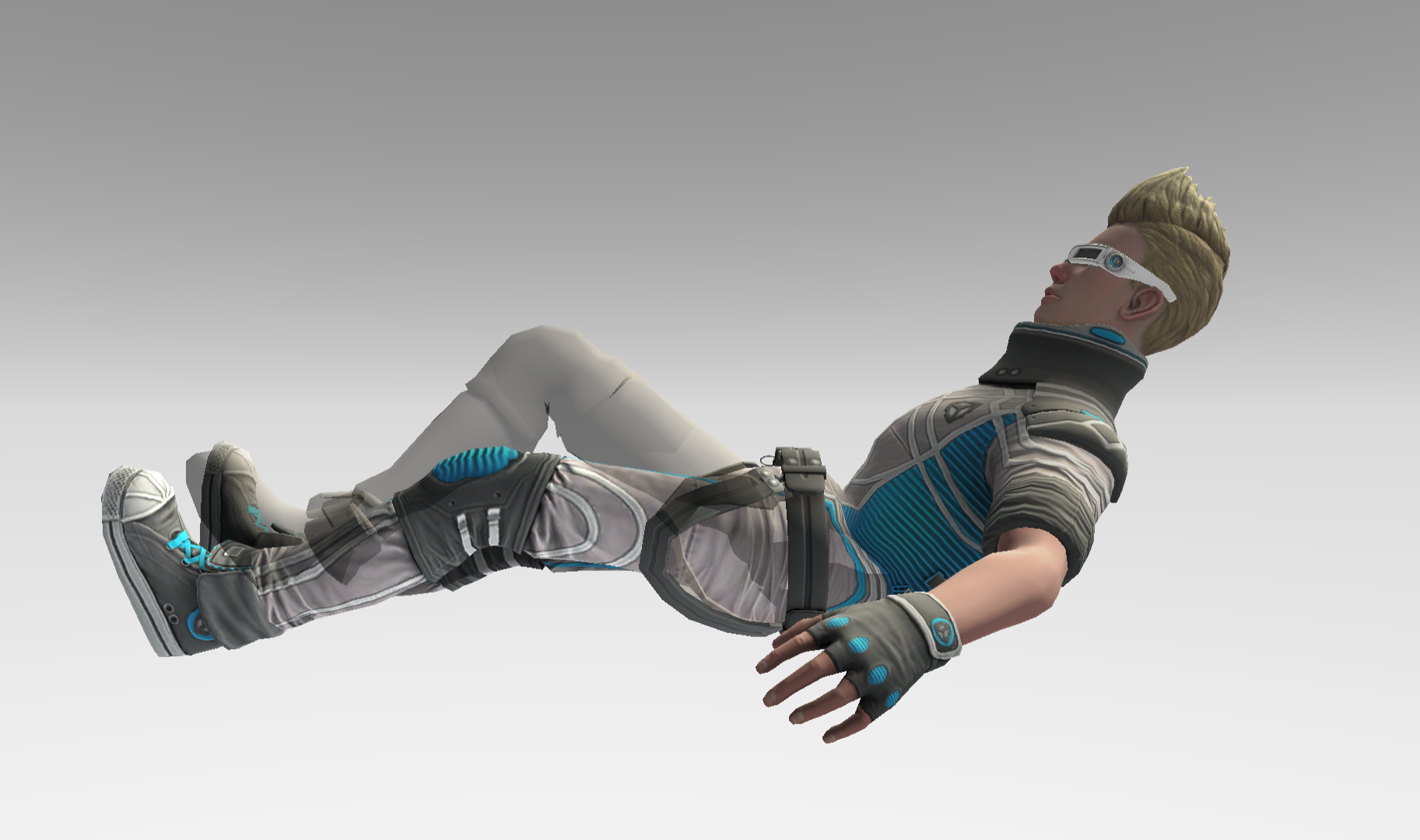}
        \caption{\textbf{TPN}: third point of view without hand clues}
        \label{fig:tpn}
     \end{subfigure}
     \hfill
     \begin{subfigure}{0.24\textwidth}
        \centering
        \includegraphics[width=\linewidth]{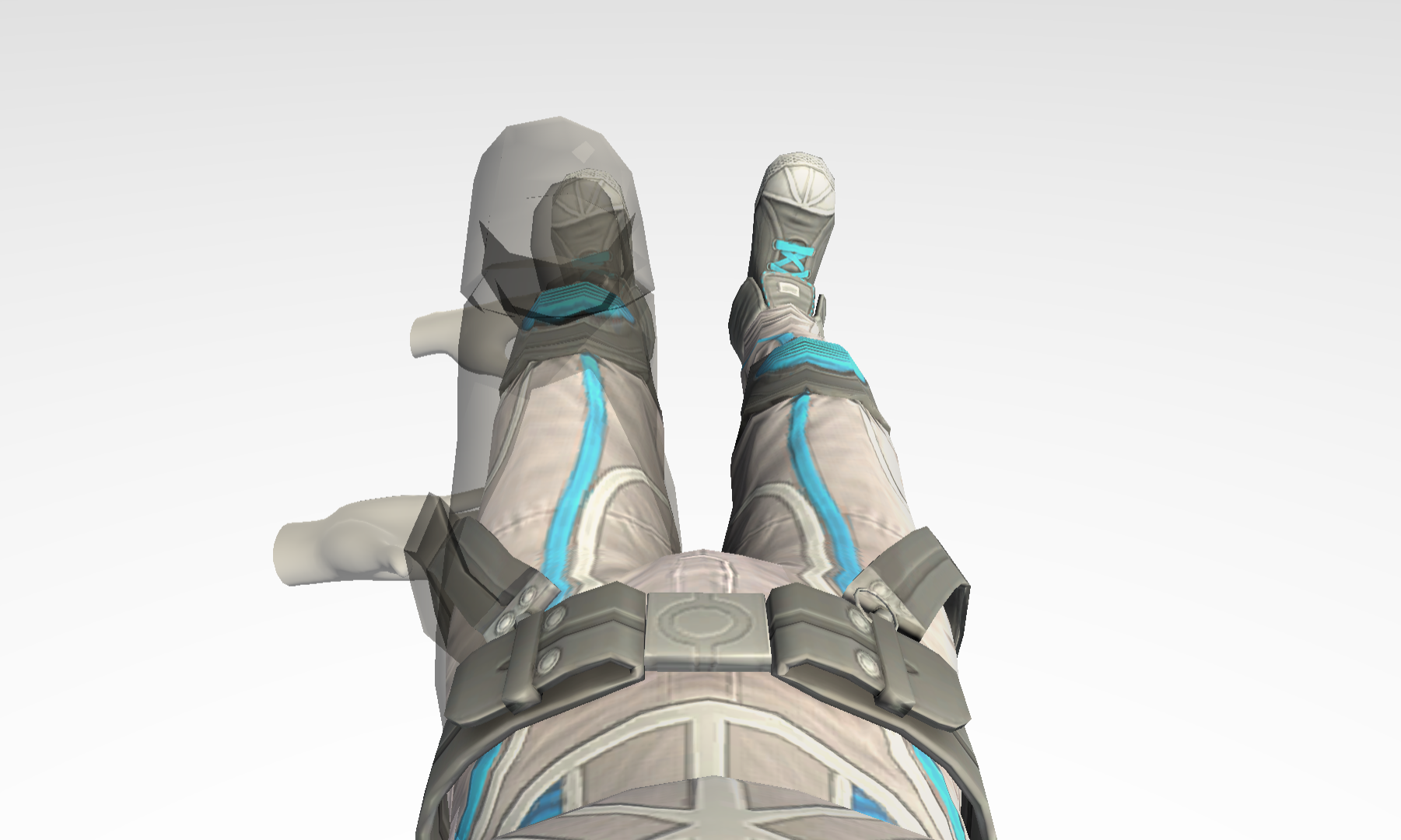}
        \caption{\textbf{FPH}: first point of view with hand clues}
        \label{fig:fph}
     \end{subfigure}
     \hfill
     \begin{subfigure}{0.24\textwidth}
        \centering
        \includegraphics[width=\linewidth]{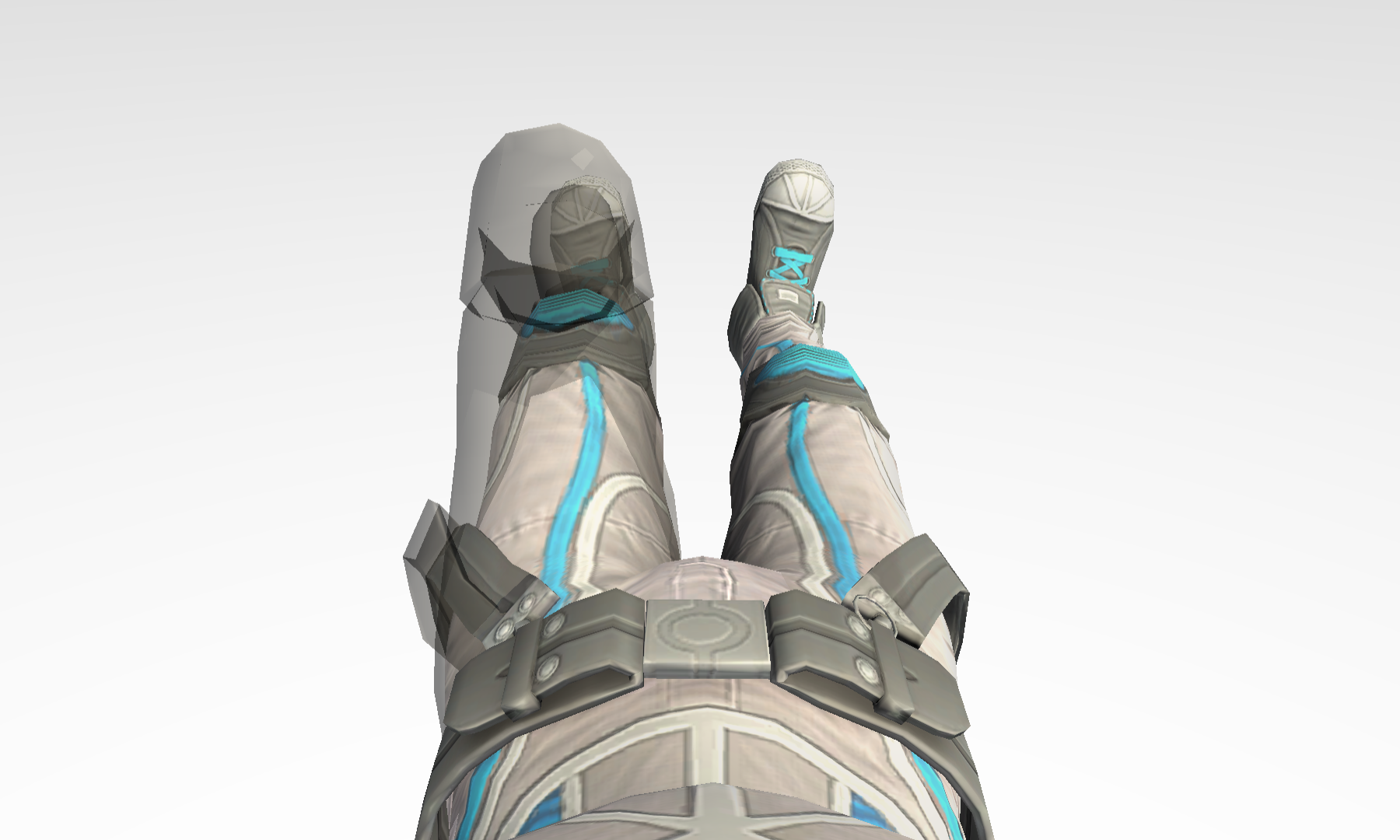}
        \caption{\textbf{FPN}: first point of view without hand clues}
        \label{fig:fpn}
     \end{subfigure}
    \vspace{-3mm}
     \caption{Illustration of four different visual feedback conditions.}
     \label{fig:visual_conditions}
\vspace{-5mm}
\end{figure}

We leveraged virtual hands to unobtrusively represent the hands of the therapists as authorities.
A pair of hands are placed around the thigh and the calf in opposite directions from the target pose to indicate the intended moving direction (Fig~\ref{fig:visual_conditions}).
We evaluated four conditions: \textbf{TPH}: third pov with virtual hands, \textbf{TPN}: third pov without virtual hands, \textbf{FPH}: first pov with virtual hands, and \textbf{FPN}: first pov without virtual hands.
Examples of all conditions for the pose \textit{bed-supported knee bends} are shown in Fig~\ref{fig:visual_conditions}. 
The target poses were kept the same as Section~\ref{observation}.
We counter-balanced the order of the conditions by a latin-square, and the order of the poses in each condition is random. 

\vspace{-3mm}
\subsubsection{Participants}
We recruited 16 participants (6 females, 10 males) from a local university with an average age of 21.5 (SD = 2.3) and an average self-reported familiarity to VR of 2.88 (SD = 1.4) out of 5 ($1$: Not familiar at all; $5$: Very familiar). 
None had perception disorders or vision problems. We tested half of the participants' left legs and the rest's right legs. Each participant received a 15 USD compensation for the 40-minute study. 

\vspace{-3mm}
\subsubsection{Apparatus}
The IMUs straps (Sec~\ref{sec:motiontracking}) were used for tracking.
We developed a VR application with Unity using the Oculus Quest 2~\footnote{https://www.oculus.com/quest-2/} headset with a desktop computer (Intel 16-Core i7 CPU@3.30 GHz, 32 GB RAM, NVIDIA GeForce RTX 3080 GPU).
The IMUs' data was transmitted to the VR application via Bluetooth. 

\vspace{-3mm}
\subsubsection{Procedure}
We conducted the user study in a lab setting. 
Participants were first informed of the study purpose and the procedure. 
After providing the demographic information, they were instructed to sit on a yoga mat.
The researchers helped the participants wore the IMU straps and the VR headset.
Participants familiarized themselves with the VR environment and the virtual avatar.
We tested three poses in a random order with 5 repetitions in each test session. 
Each participant thus performed $4$ conditions/sessions $\times$ $3$ poses $\times$ 5 repetitions $= 60$ movements.

After each session, the participant was asked to grade the following five statements using a 5-point Likert scale (1: strongly disagree, 3:neutral, 5: strongly agree): \textit{"I could quickly understand the visual instructions"} (information presentation), \textit{"I was well aware of my motions"} (self-perception), \textit{"I felt motivated"} (motivation), \textit{"I felt stressed"} (stress), and \textit{"I liked exercising with the system"} (overall impression).
Users took 5-minute breaks between sessions. 
After all the sessions were completed, participants could revise and modify their scores for the conditions. Lastly, we conducted a 5-minute interview to investigate their opinions on the visual feedback. 

\vspace{-3mm}
\subsubsection{Results}
As mentioned in Section~\ref{sec:automated}, we evaluated the following objective metrics.

\begin{enumerate}
    \item \textbf{Understanding time} is the duration of the understanding phase when they see a pose for the first time in a session. This indicates the time taken to understand new instructions. 
    \item \textbf{Angle deviation} is the angular difference between the starting pose of the holding phase and the target. This metric measures how accurately participants can achieve the target pose with the system.
    \item \textbf{Effective time} is the duration within the 10-second holding phase in which the participant stably holds the starting pose without large-scale fluctuations. We defined a threshold of 4-degree deviation ($\pm 2$ degrees) within which the holding is stable and effective. This metric is indicative of the stability of the holding phase. 
\end{enumerate}

The variables for the objective metrics did not follow a normal distribution ($p < 0.05$ under Shapiro–Wilk Test for all tested variables).
We thus used Wilcoxon signed-rank test ($p < 0.05$) for statistic analysis on the objective metrics and the subjective evaluation.

Statistical analysis results revealed a significant effect of the \textbf{pov} on the \textit{understanding time} $(Z = 7.03,~p < 0.001)$. 
This indicates that users take much less time to understand the instructions using the third pov in VR. 
Table~\ref{tab:visual_results} shows no significant effect of \textbf{the presence of authority} on the objective metrics. 
As shown in Table~\ref{tab:visual_results}, statistical analysis results indicated that the third pov significantly outperforms the first pov on \textit{information presentation} ($Z = 2.69,~p < 0.01$), \textit{self perception} ($Z = 4.12,~p < 0.001$), \textit{stress} ($Z = 2.15,~p = 0.032$), and \textit{overall impression} ($Z = 2.92,~p < 0.01$). However, there is no statistical significant effect of \textbf{presence of authority} on the qualitative metrics. 

14 out of 16 participants indicated in the interviews that they found the third pov more suitable for physical rehabilitation poses as it offers a clearer and more complete view of their bodies without demanding extra physical effort, especially for the pose prone leg raises.
In summary, results revealed that \systemname should adopt the third pov without the virtual hand.

\begin{table}
  \begin{threeparttable}
  \begin{tabular}{cccccc}
    \toprule
    Metrics & TPH & TPN & FPH & FPN & p-value\tnote{a} \\ 
    \midrule
    Angle deviation(degree) & $4.65 \pm 1.94$ & $4.66 \pm 1.80$ & $5.48 \pm 2.80$ & $5.03 \pm 2.19$ & NS \\
    Effective time(s) & $7.01 \pm 3.10$ & $6.92 \pm 3.05$ & $7.09 \pm 3.24$ & $7.14 \pm 2.93$ & NS \\
    Understanding time(s) & $9.50 \pm 8.40$ & $8.03 \pm 5.50$ & $24.07 \pm 19.93$ & $24.81 \pm 24.33$ & p < 0.001 \\
    Information presentation & $3.31 \pm 0.99$ & $3.15 \pm 0.77$ & $2.62 \pm 0.62$ & $2.54 \pm 0.50$ & p < 0.01 \\
    Self perception & $4.77 \pm 0.42$ & $4.77 \pm 0.42$ & $4.00 \pm 0.78$ & $3.85 \pm 0.36$ & p < 0.001 \\
    Movtivation & $4.00 \pm 0.78$ & $3.69 \pm 0.46$ & $3.46 \pm 0.50$ & $3.38 \pm 0.92$ & NS \\
    Stress & $3.92 \pm 1.14$ & $4.23 \pm 0.42$ & $3.54 \pm 0.53$ & $3.54 \pm 1.15$ & p = 0.032 \\
    Overall & $3.69 \pm 0.46$ & $3.92 \pm 0.47$ & $3.31 \pm 0.72$ & $2.77 \pm 0.80$ & p < 0.01 \\
    \bottomrule
  \end{tabular}
  \begin{tablenotes}
    \footnotesize
    \item Data are summarized as mean $\pm$ standard deviation. NS indicates no significance.
    \item[a] Comparison of point of view. \textbf{TPH} and \textbf{TPN} vs. \textbf{FPH} and \textbf{FPN}.
  \end{tablenotes}
  \end{threeparttable}
  \caption{Objective and subjective results of the visual feedback characterization study.}
  \label{tab:visual_results}
\vspace{-10mm}
\end{table}

\vspace{-3mm}
\subsection{Haptic Feedback Characterization}
\label{sec:haptic}
We further characterized \systemname's the haptic feedback by evaluating different haptic methods' effectiveness in providing motion guidance. 
Sec~\ref{sec:unimodal_literature} and Sec~\ref{sec:multimodal_literature} discussed different haptic feedback methods that have been leveraged for motion guidance.
Skin stretch displays on legs usually require setting up exoskeleton devices and EMS induces involuntary movement, making them unfit for TKA rehabilitation. 
We thus investigate the feasibility of using vibrotactile and pneumatic feedback to provide motion guidance compared to a no haptic feedback baseline.

\vspace{-3mm}
\subsubsection{Design}
We evaluated three conditions: \textbf{None}: no haptic feedback, \textbf{Vibro}: vibrotactile feedback, and \textbf{Pneu}: pneumatic feedback. 
All visual display setup and other study designs were kept the same as Sec~\ref{sec:visual}.

\vspace{-3mm}
\subsubsection{Participants}
We recruited 11 participants (5 females, 6 males) from a local university with an average age of 24.1 ($SD = 4.0$) and an average self-reported familiarity to VR of 2.18 ($SD = 1.0$) out of 5 ($1$: Not familiar at all; $5$: Very familiar).
None had perception disorders or vision problems.
We tested 6 of the participants' left legs and the rest's right legs.
Each participant received a 15 USD compensation for the 40-minute study.

\vspace{-3mm}
\subsubsection{Apparatus}
The hardware setup of the visual system was the same as Sec~\ref{sec:visual}. 
We prototyped straps to be worn at the thigh and the calf for vibrotactile/pneumatic feedback.
Feedback can be actuated at the up and down sides of the strap to provide directional information.
The vibrotactile actuators(9000 rpm) were screwed directly onto the straps.
A total of four vibrators for one vibrotactile haptic system were powered and controlled by an Arduino Uno connected to the computer. 
We adopted the fabrication method for prototyping inflatable structures in aeroMorph~\cite{ou_16}.
Airbags of 7cm wide and 8cm long were glued to the inner side of the straps so the airbags were in contact with the skin (exerts a ~10N force when inflated).
Each airbag was controlled by two micro-pumps (12 V, 5 W) for inflation and deflation. 
A total of 8 micro-pumps were controlled by an 8-way relay and an Arduino Mega~\footnote{https://www.arduino.cc/en/Main/arduinoBoardMega2560} connected to the computer. 
The micro-pumps were powered by a regulated power supply and a step-down converter. 

\vspace{-3mm}
\subsubsection{Procedure}
The study procedure was the same as in Sec~\ref{sec:visual}. 
With the help of researchers, the participants wore IMU straps, vibrotactile/pneumatic feedback straps, and the VR headset.
We tested three poses, straight leg raises, prone straight leg raises, and bed-supported knee bends Fig.~\ref{fig:poses}) with 5 repetitions in each test session. Therefore, each participant performed $3$ conditions/sessions $\times$ $3$ poses $\times$ 5 repetitions $= 45$ movements in this user study.

\begin{table}
  \begin{threeparttable}
  \begin{tabular}{ccccccc}
    \toprule
    Metrics & None & Vibro & Pneumatic & p-value\tnote{a} & p-value\tnote{b} & p-value\tnote{c} \\ 
    \midrule
    Angle deviation(degree) & $5.11 \pm 1.28$ & $0.39 \pm 0.38$ & $0.30 \pm 0.39$ & p < 0.001 & p < 0.001 & NS \\
    Effective time(s) & $6.92 \pm 4.42$ & $8.73 \pm 3.24$ & $8.17 \pm 3.45$ & p < 0.001 & p < 0.01 & NS \\
    Understanding time(s) & $8.81 \pm 6.31$ & $12.87 \pm 8.29$ & $13.66 \pm 9.38$ & p = 0.036 & p = 0.021 & NS \\
    Information presentation & $3.64 \pm 1.07$ & $4.55 \pm 0.66$ & $3.82 \pm 0.94$ & p = 0.025 & NS & p = 0.023 \\
    Self perception & $4.55 \pm 1.16$ & $4.82 \pm 0.39$ & $4.16 \pm 1.11$ & NS & p = 0.042 & p = 0.034 \\
    Movtivation & $4.09 \pm 0.79$ & $4.64 \pm 0.45$ & $4.45 \pm 0.66$ & p = 0.034 & NS & NS \\
    Stress & $4.18 \pm 1.11$ & $4.91 \pm 0.29$ & $4.55 \pm 0.66$ & p = 0.046 & NS & NS \\
    Overall & $4.09 \pm 1.08$ & $4.82 \pm 0.39$ & $4.55 \pm 0.66$ & p = 0.038 & p = 0.044 & NS \\
    \bottomrule
  \end{tabular}
  \begin{tablenotes}
    \footnotesize
    \item Data are summarized as mean $\pm$ standard deviation. NS indicates no significant effect.
    \item[a] Comparison of \textbf{None} vs. \textbf{Vibro}.
    \item[b] Comparison of \textbf{None} vs. \textbf{Pneumatic}.
    \item[c] Comparison of \textbf{Vibro} vs. \textbf{Pneumatic}.
  \end{tablenotes}
  \end{threeparttable}
  \caption{Objective and subjective results of the haptic feedback characterization study.}
  \label{tab:haptic_results}
\vspace{-10mm}
\end{table}

\vspace{-3mm}
\subsubsection{Results}
We adopted the same metrics and the statistical analysis methods as in Sec~\ref{sec:visual}.
Statistical analysis results revealed a significant effect on all objective metrics as shown in Table~\ref{tab:haptic_results} $(p < 0.05)$. 
The increased angle deviation in \textbf{None}, compared to \textbf{Vibro} $(Z = 5.40,~p < 0.001)$ and \textbf{Pneu} $(Z = 6.19,~p < 0.001)$ verifies the supplemental haptic feedback' effect in enhancing performance.
Both \textbf{Vibro} $(Z = 4.04,~p < 0.001)$ and \textbf{Pneu} $(Z = 2.64,~p < 0.01)$ outperformed \textbf{None} in helping users remain stable during the holding, shown by their longer effective time.
Both \textbf{Vibro} $(Z = 2.10,~p = 0.036)$ and \textbf{Pneu} $(Z = 2.30,~p = 0.021)$ had longer understanding time due to the time needed for actuation and for processing the haptic guidance.
Table~\ref{tab:haptic_results} shows no significant effect of other factors. 


The subjective results showed a significant effect of the haptic feedback method on all subjective metrics, as shown in Table~\ref{tab:haptic_results}.
\textbf{Vibro} was rated higher on information presentation compared to \textbf{None} $(Z = 2.23,~p = 0.025)$ and \textbf{Pneu} $(Z = 2.27,~p = 0.023)$.
Both \textbf{Vibro} $(Z = 2.12,~p = 0.034)$ and \textbf{None} $(Z = 2.03,~p = 0.042)$ were better on self perception compared to \textbf{Pneu}.
Participants explained in the interview that the intense vibrations conveyed clear instructions, whereas the pneumatic feedback is slower and more subtle.
This created some confusion for the participants when using pneumatic feedback to convey rapidly fluctuations during the holding phase.
Results revealed that compared to \textbf{None}, participants were more motivated $(Z = 2.12,~p = 0.034)$ and also more stressed $(Z = 2.00,~p = 0.046)$ with \textbf{Vibro}. 
This was due to intense and numbing vibrations throughout the reaching phase and participants' knowledge that fluctuations during the holding phase would trigger vibration "warnings".
Overall, participants preferred \textbf{Vibro} $(Z = 2.07,~p = 0.038)$ and \textbf{Pneu} $(Z = 2.00,~p = 0.044)$ over \textbf{None} while exercising.
This indicates that despite a mildly increased understanding time, participants found haptic feedback helpful in providing motion guidance and awareness.

In the interview, half of the participants preferred vibrotactile due to its intense and instantaneous response, while airbags took longer to inflate.
The rest liked pneumatic feedback's soothing and pseudo-kinesthetic sensation.

\begin{figure}[htbp]
    \begin{minipage}{0.5\textwidth}
     \centering
     \begin{subfigure}{0.3\textwidth}
        \centering
        \includegraphics[width=\linewidth]{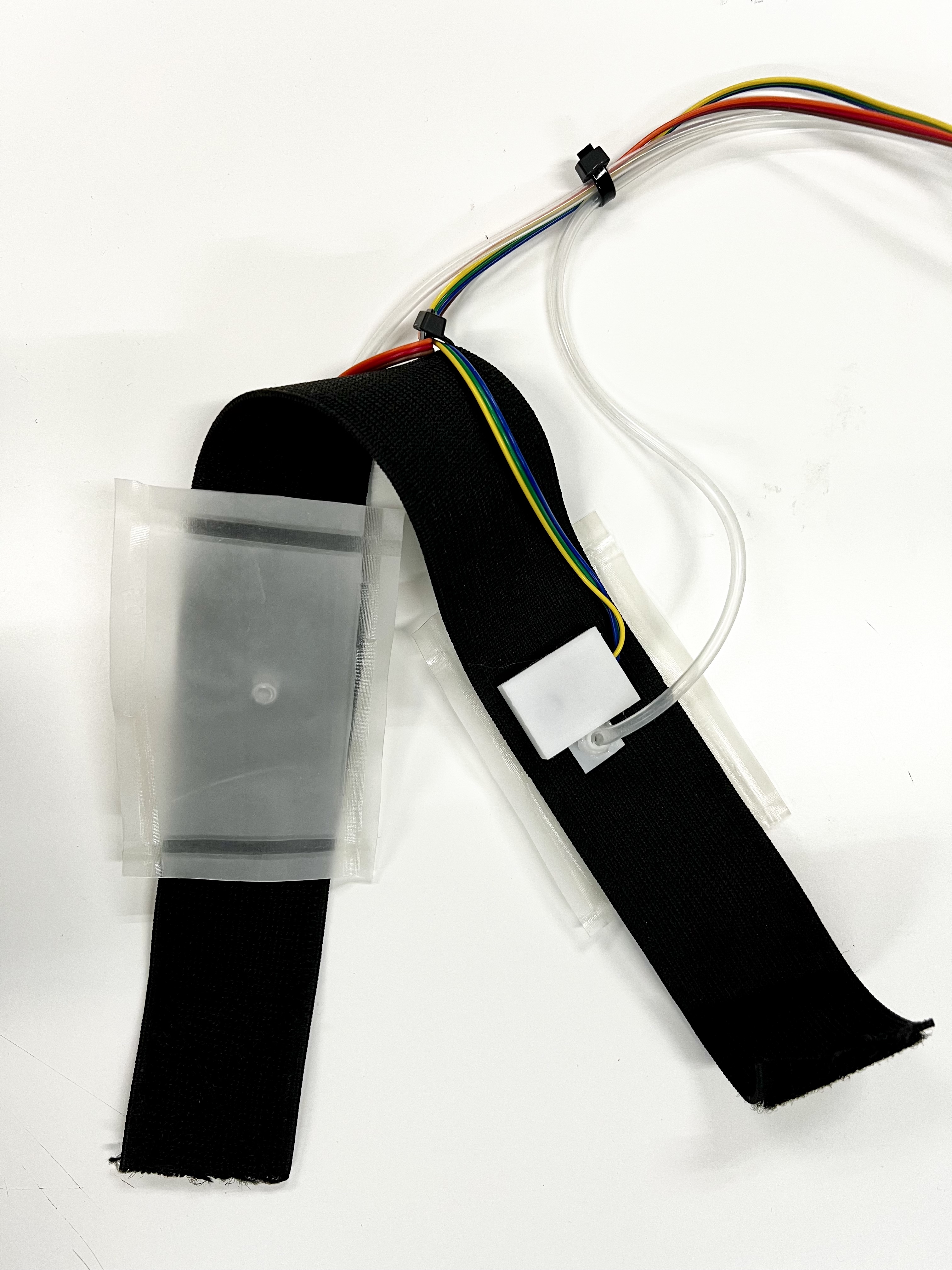}
        \caption{ }
        \label{fig:hapticcombine}
     \end{subfigure}
     \hspace{0.01em}
     \begin{subfigure}{0.66\textwidth}
        \centering
        \includegraphics[width=\linewidth]{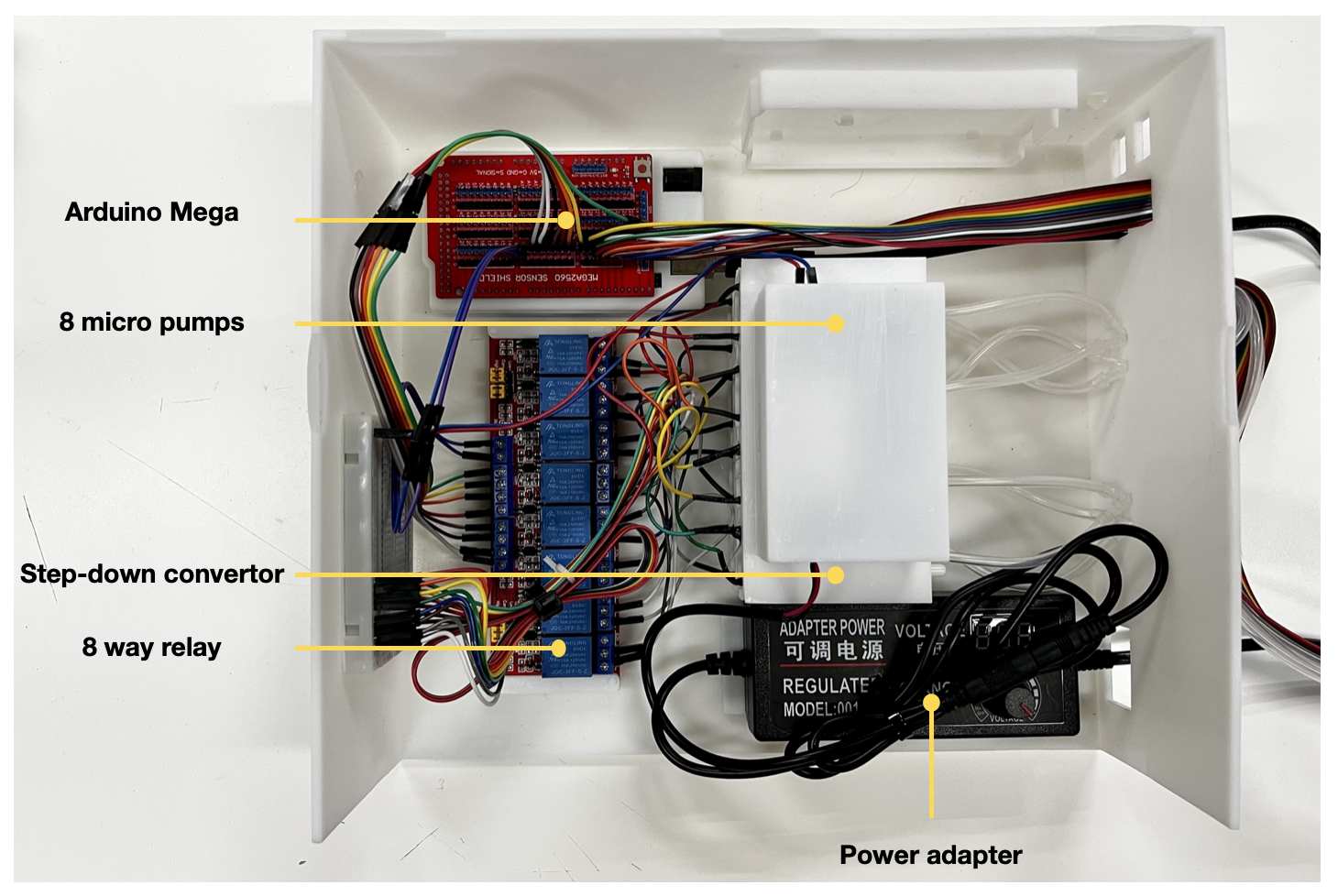}
        \caption{ }
        \label{fig:implementation}
     \end{subfigure}
     \vspace{-3mm}
     \caption{(a): We assemble each pair of vibrators and air compartments on a strap. (b): The entire haptic system in a 3D printed box.}
    \end{minipage}
    \hspace{0.05em}
    \begin{minipage}{0.45\textwidth}
     \begin{subfigure}{0.47\textwidth}
        \centering
        \includegraphics[width=\linewidth]{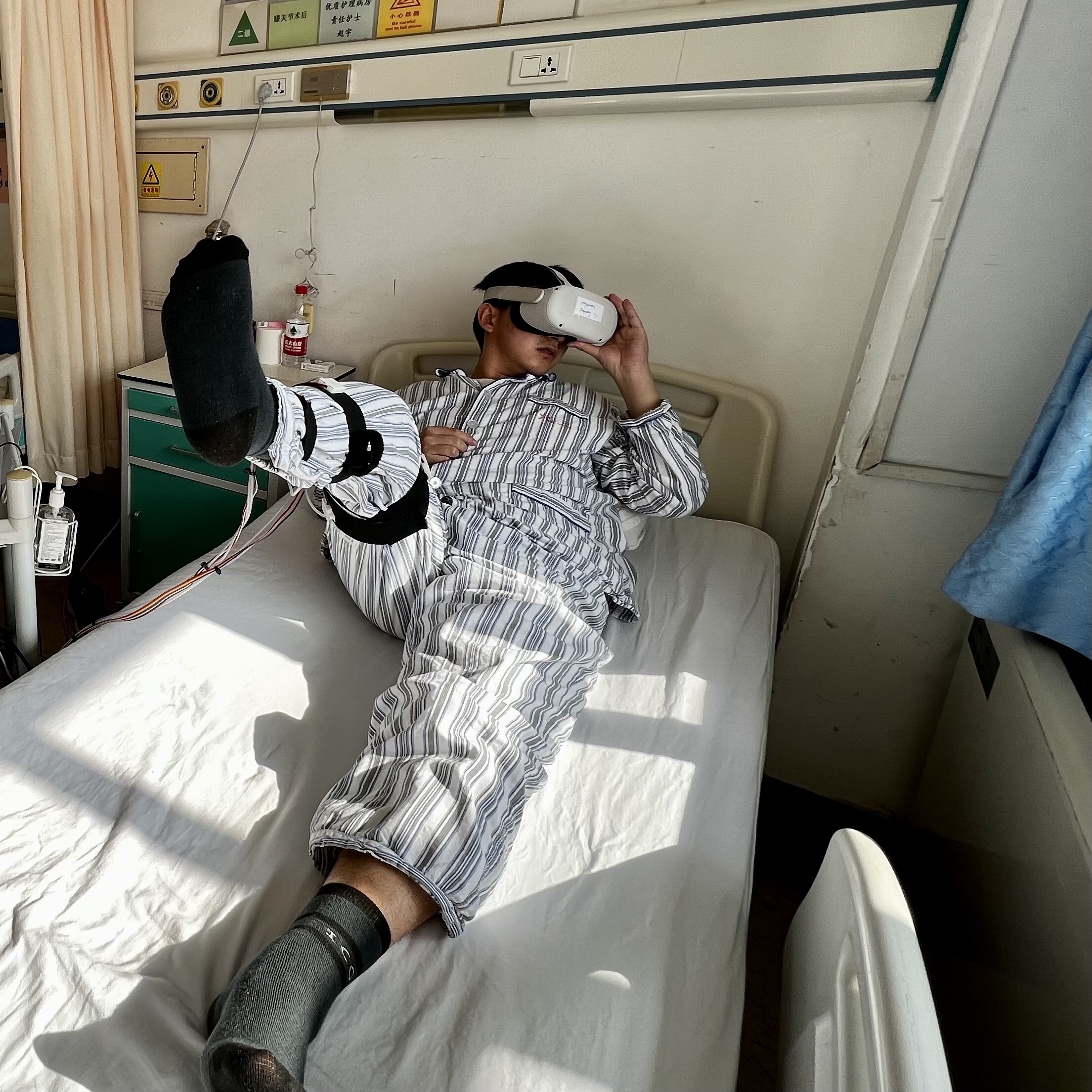}
        \caption{}
        \label{fig:evaproa}
     \end{subfigure}
     \hspace{0.02em}
     \begin{subfigure}{0.47\textwidth}
        \centering
        \includegraphics[width=\linewidth]{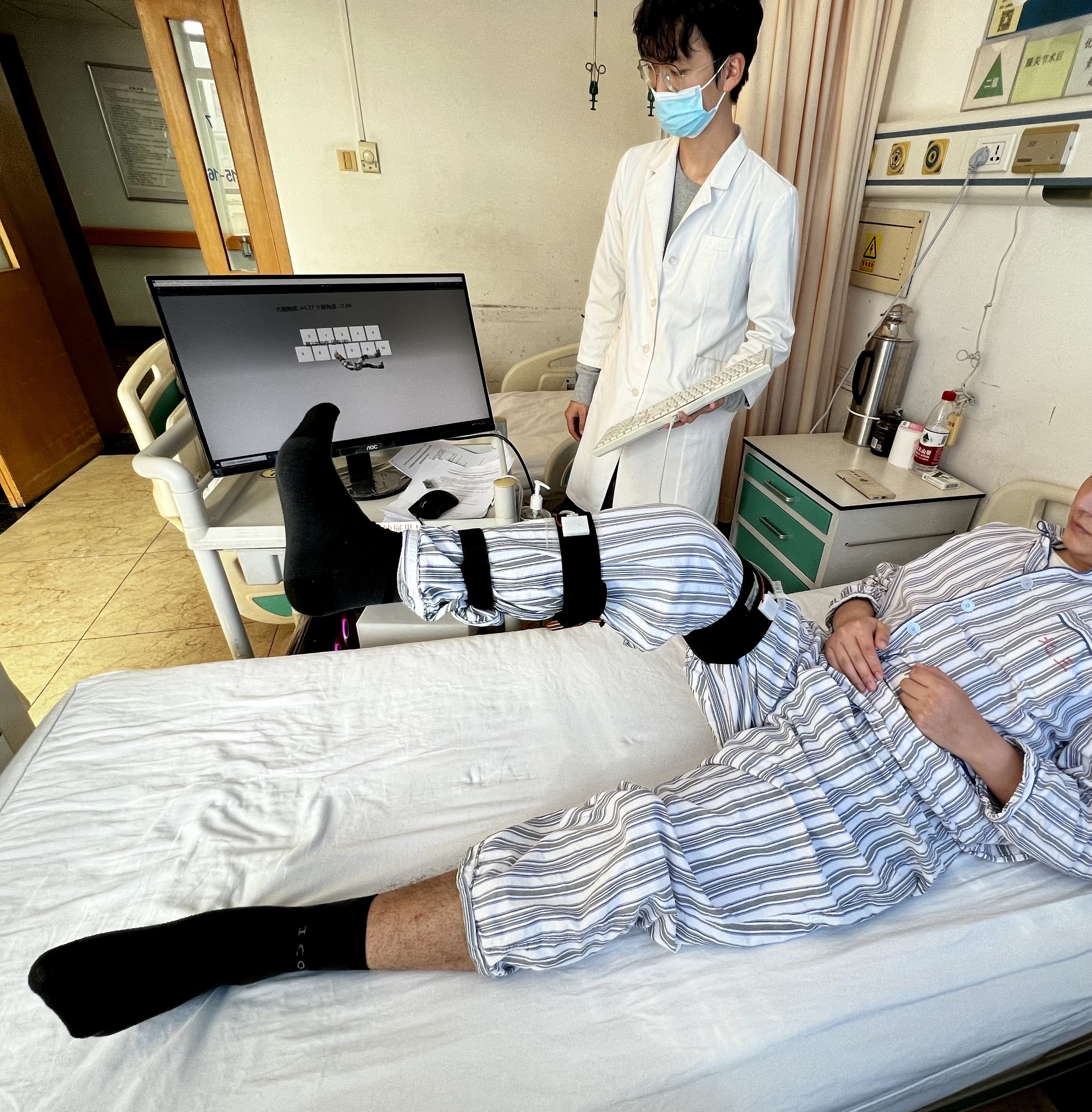}
        \caption{}
        \label{fig:evaprob}
     \end{subfigure}
     \vspace{-3mm}
     \caption{(a) Users complete the exercises with \textbf{\systemname}. (b) Users complete the exercises with \textbf{SD}.}
    \label{sec:evaprocedure}
    \end{minipage}
\end{figure}

\vspace{-3mm}
\subsubsection{Finalizing haptic feedback}
Based on the study results, we combined vibrotactile and pneumatic feedback's respective advantages in our implementation of \systemname. 
Specifically, we mapped these two haptic feedback methods to different phases of a movement. 
Pneumatic feedback is provided during reaching. 
The gradually exerted pressure on the skin notifies the user to slowly and safely reach the target pose.
Once \systemname detects that the user has stabilized and starts the holding phase, vibrotactile feedback is provided to supervise the user during holding.
The corresponding vibrator is actuated when the user goes below or above the acceptable holding range.
Upon receiving instantaneous and intense "warnings," the user can timely adjust his/her pose.

We designed and 3D-printed cases to assemble each pair of vibrators and airbags and mounted the assembly onto a strap (Fig~\ref{fig:hapticcombine}).
The vibrators and the relay, which control the pneumatic system, are altogether controlled by an Arduino Mega connected to a computer through a serial port.
We 3D printed a box (278mm in length, 246mm in width, and 98mm in height) to house all parts compactly so that \systemname can be easily mobilized.
Fig~\ref{fig:implementation} demonstrates the assembled box.

\vspace{-3mm}
\section{Clinical Evaluation}
\label{sec:evaluation}


\subsection{Design}
We compared \textbf{\systemname} to \textbf{SD}, a screen display based system without haptic guidance.
For \textbf{SD}, participants wear only the tracking module to complete the same exercise with the same visual guidance as in \textbf{\systemname} but displayed on a screen and without haptic feedback.
We co-designed the study with 3 therapists and were advised that the study should take $15$-$20$ minutes in total to reduce the amount of exercise to prevent excessive exercises.
We replaced \textbf{prone straight leg raises} (muscle strength exercise) with \textbf{knee extensions} (range of motion exercise) compared to characterization studies in Sec~\ref{sec:system_design} to reduce the physical effort needed. 
Thus each participant was asked to perform 2 repetitions of \textbf{straight leg raises}, \textbf{bed-supported knee bends}, and \textbf{knee extensions} (Fig.~\ref{fig:poses}) in a session.
The objective metrics remained the same with the characterization study in Sec~\ref{sec:system_design}.
We also conducted semi-structured interviews to further investigate the participants' responses to the usability of \systemname and their willingness to use \systemname for long-term at-home physical rehabilitation.
This study was approved by the hospital's IRB board.

\vspace{-3mm}
\subsection{Participants and apparatus}
We recruited 10 participants (referred to as P1...P10), 6 females and 4 males with an average age of 63.2($SD=15.20$), from the local hospital through convenience sampling.
The time since their surgery was 2-7 days ($M=4.60, SD=1.43$) and they were suggested by the therapists to be in a similar stage in recovery.
3 of them had the surgery on their left legs, while 7 on their right legs.
None of them had experience with VR, perception disorders, or vision problems.

The implementation of \systemname has been elaborated in Sec~\ref{sec:system_design}.
For condition \textbf{SD} we removed the haptic feedback wearables and used a AOC G2490VX 24" monitor screen (1920x1080 resolution and 144Hz refresh frequency) set up beside the participants' beds to display the VR scene. 
The entire system was set up on a mobile cart.

\vspace{-3mm}
\subsection{Procedure}
We firstly explained the purpose and the procedure of the study and asked the participant to sign an agreement and fill in their demographic information.
The participants' therapists-in-charge set the target angle for each posture based on which we set the target pose in the system.
Then a researcher helped them wear the required wearables, including two IMU straps, a VR headset, and two haptic feedback straps if using \textbf{\systemname}. 
After ensuring that the patients were comfortable and the wearables did not obstruct their movement, the researcher showed a sample scene in VR or on the screen to familiarize the patients with the visual displays.
If in \textbf{\systemname}, a sample haptic feedback for a complete movement was also demonstrated and the feedback mechanism explained.
We counterbalanced the conditions for each participant.
Similar to the study procedure in Sec~\ref{sec:system_design}, we tested three poses with 2 repetitions in each test session and 5-minutes breaks between sessions.
Therefore, each participant performed $2$ conditions/sessions $\times$ $3$ exercise poses $\times$ $2$ repetitions $= 12$ movements in this evaluation. 
During the session, at least one therapist was present to guarantee the safety of the sessions and to provide his/her input for the systems.
Then we conducted a semi-structured interview with the participant, which lasted for about 20 minutes.
Two researchers conducted the interviews and took turns to ask questions.
All interviews were audio-recorded with permission and transcribed.

We began the interview with "ice breaker" surveys  \textit{Q1: How's your recovery?} and \textit{Q2: Do you feel alright?} to understand their recovery status.
We asked their opinion about the visual motion guidance's effectiveness and clarity: \textit{Q3: Do you think the target pose helped you complete the exercise?} and \textit{Q4: Did you have trouble understanding the visual guidance?}.
We then asked about the automated progress tracking in the systems: \textit{Q5: Do you think the automated timing and counting help you complete the practice?}.
To evaluate the differences between \textbf{\systemname} and \textbf{SD}, we first investigated whether HMD based VR improved the patients' rehabilitation experience by probing \textit{Q6: Did the VR device help you exercise?} and \textit{Q7: Did you feel comfortable with the VR device?}.
Then we subjectively evaluated how patients reacted to the haptic feedback by asking \textit{Q8: Do you think the haptic guidance helped you complete the exercise?}, \textit{Q9: Did you have trouble understanding the haptic guidance?}, \textit{Q10: Did the haptic guidance feel comfortable?}
We also asked the patients to compare both systems to the traditional take-home document (\textit{Q11/Q12: What is the difference between \textbf{\systemname} / \textbf{SD} and the traditional guiding document?}).
Lastly, to assess \systemname's potential to facilitate self-monitored physical rehabilitation in the future, we asked \textit{Q13: Are you willing to use \systemname in the future?}, \textit{Q14: Do you believe that it could facilitate self-monitored physical rehabilitation at home?}, and \textit{Q15: Do you have any questions or advice about \systemname?}.





\vspace{-3mm}
\subsection{Results}


\begin{table}
  \begin{threeparttable}
  \begin{tabular}{cccc}
    \toprule
    Metrics & \systemname & SD & p-value\tnote{a} \\
    \midrule
    Angle deviation(degree) & $2.04 \pm 0.85$ & $1.99 \pm 0.71$ & NS \\
    Effective time(s) & $6.59 \pm 1.84$ & $6.03 \pm 1.06$ & p = 0.013 \\
    Understanding time(s) & $17.87 \pm 9.96$ & $13.90 \pm 11.31$ & p = 0.012 \\
    \bottomrule
  \end{tabular}
  \begin{tablenotes}
    \footnotesize
    \item Data are summarized as mean $\pm$ standard deviation. NS indicates no significant effect.
    \item[a] Comparison of \textbf{\systemname} vs. \textbf{SD}.
  \end{tablenotes}
  \end{threeparttable}
  \caption{Objective results of the clinical evaluation study.}
  \label{tab:eva_results}
\vspace{-12mm}
\end{table}

\vspace{-2mm}
\subsubsection{Objective Results}

We used the Wilcoxon test to analyze the objective data as the variables did not follow a normal distribution ($p < 0.05$ under Shapiro–Wilk Test for all tested variables). The results are shown in Table~\ref{tab:eva_results}.  
The objective results showed a significant effect on understanding time and effective time $(Z = 2.39,~p = 0.012)$.
\textbf{Patients took a longer time to understand the guidance with \textbf{\systemname}.}
This is in accordance with our findings in Sec~\ref{sec:haptic} and is due to the extra time needed to actuate the haptic feedback.
Users also needed a longer time to process haptic motion guidance as compared to purely visual ones in \textbf{SD}.
The effective holding time with \textbf{\systemname} is also longer $(Z = 2.48,~p = 0.013)$.
\textbf{Patients were thus able to hold their poses with improved stability with vibrotactile warnings and a VR environment. }
As angle deviation didn't exhibit significant differences between \textbf{\systemname} and \textbf{SD}, \textbf{this indicates that the patients' range of motion capability remains the same in this short-term evaluation.}

\vspace{-3mm}
\subsubsection{Interview Analysis}
Both researchers separately transcribed the recording and reviewed and corrected the transcripts.
We first created a set of codes related to different aspects of the system.
We then developed a codebook with these initial codes and coded the interview transcripts iteratively.
After each iteration, the two coders discussed their findings and new codes.
We summarize and discuss our findings below.

\textbf{Motion awareness and guidance is effectively delivered through motion reconstruction and the overlaid target avatar.}
7 out of 10 participants suggested that the major difference between \systemname and the traditional document was the motion reconstruction and the target avatar, which was found clear and helpful.
They were interested in observing their reconstructed motion visually since they can \textit{"clearly see how they are moving in real-time which help a lot in grasping what they are lacking and where their legs need to go"}(P6).
7 participants thought the target avatar helpful as they didn't have to \textit{"look back and forth to compare their motion and the target"} (P5), and the visual discrepancy between the reconstructed leg and the target is \textit{"...intuitive and clear"} (P8).
One participant pointed out the confusion caused by the white virtual background and preferred that \textit{"the scene is his home and the avatar looks like himself"} (P3).

\textbf{Automated progress tracking is effortless and supervisory. }
All participants liked the automated timing and movement counting, which saved them from manually tracking the progress themselves and helped them to concentrate.
P1 commented that \textit{"I think I might practice a few seconds less than the required if I was timing on my own. I can also be more attentive to my motion and the target pose"}.
The automation also provides supervision for malingering patients.
One participant complained that \textit{"I felt supervised to complete compulsive tasks since I couldn't quit midway"} (P5).


\textbf{Elder patients liked the immersion in VR but hesitated to adopt new technologies.}
None of the participants reported discomfort or dizziness when seeing the VR scenes.
Half of the participants showed interest in the HMDs and liked that VR was able to construct a virtual surrounding which helped avoid distractions.
One participant commented that \textit{"...the HMD blocked out the real world so I couldn't see messy environment or other people around me."}(P5) 
However, despite these recognized advantages of VR over screen displays, the rest of the participants indicated that screen displays were \textit{"good enough"}(P6).
We noticed that the participants, who were mostly elderly people, had concerns about the cost and difficulty of setting up the VR devices such that they hesitated to adopt HMDs in their daily life.
P4 said that \textit{"neither my son nor I have ever used this thing(HMD), is it hard to learn?"} while P7 also mentioned that \textit{"I like the virtual scene....but I do not want a new device in my home...my grandsons would take it away and play games..."}.

\textbf{Patients found haptic feedback to be subtle but helpful. }
None of the participants reported discomfort with both haptic feedback methods.
Only 6 out of the 10 perceived the haptic feedback, and among them, 3 thought haptic feedback to be helpful.
Patients also reported that pneumatic feedback is harder to notice than vibrotactile ones.
On this, the therapists commented that the patient's legs were less sensitive and responsive to an external stimulus due to the surgery.
Those who perceived the haptic feedback thought the warnings conveyed by vibrations helpful as one participant said \textit{"I could feel something on my leg if it started to fall...I rose my leg up as soon as I felt the vibration and then it stopped"}(P9).
Some participants also reported that though they perceived the haptic feedback, they were \textit{"too occupied with the visual displays"} (P10) such that they ignored or didn't understand what the haptic feedback meant.




\textbf{Patients were willing to use \systemname when a clincal session is not an option. }
8 out of the 10 participants were willing to use \systemname for their long-term physical rehabilitation at home and 7 were confident that \systemname could facilitate their training.
Participants also conveyed their preference for exercising with therapists.
P6 commented that \textit{"I am willing to use this after discharge as long as it is not very expensive...but I don't think I will use it in the hospital since the therapist can guide and help me."}
3 participants expressed fear in the hospital replacing therapist-supervised sessions with computer-assisted system.

\vspace{-8mm}
\section{Discussion}

\vspace{-1mm}
\subsection{Different visualizations of VR environments and avatars}
Previous works showed that the environment affects and can potentially help the recovery of patients~\cite{Grassini_20}.
Similarly, in our clinical evaluation study (Sec~\ref{sec:evaluation}), patients suggested changing the VR scene to a familiar home setting and letting the virtual avatars adopt the patients' appearances to enhance embodiment.
Yet to avoid any possible distraction or confusion caused by the visualized environments and to prioritize the system's capability to facilitate self-monitored physical rehabilitation, we used simplistic avatars and background to effectively provide motion awareness and motion guidance.
In future work, we will investigate how different visualizations might help further improve users' performance.


\vspace{-2mm}
\subsection{Designing asynchronous involvement of the therapists}
Our implementation \systemname was a partial implementation of the design recommendations presented in Sec~\ref{sec:interview}.
\systemname focused on optimizing the patient's physical rehabilitation experience and enhancing the effectiveness of sessions. We leave investigating the asynchronous involvement of the therapists in a self-monitored physical rehabilitation system to future work.
In our clinical evaluation (Sec~\ref{sec:evaluation}), the asynchronous involvement was realized by therapists informing the researchers about the physical rehabilitation goals before each patient's session which can be automated
Involvement of the therapists would also require investigation into what information therapists need to know and in what form to achieve effective asynchronous information delivery. 

\vspace{-2mm}
\subsection{Limitation and future work}
In \systemname the tracking data, VR visual displays, and haptic feedback controls were synchronized and controlled via a computer for the ease of researchers closely monitoring the entire system during studies.
The use of a computer as a central controller, however, is not necessary and can be replaced by a HMD in the future.

In the evaluation study~\ref{sec:evaluation}, we conducted a short-term clinical study with TKA patients with a 20 minutes exercising session for each patient. 
We recognize that the goal of self-monitored physical rehabilitation is to guarantee the quality of every session that the patients carry out to finally result in long-term improvements.
However, we were not able to carry out long-term studies due to IRB requirements and the difficulty of recruiting patients after their discharge given the changing COVID-19 situations.
We validated the usability and potential to effectively facilitate self-monitored physical rehabilitation through our short-term evaluation studies. We plan to further evaluate \systemname with long-term studies in the future when the situation permits.

In our design recommendations presented in Section~\ref{sec:interview}, we discussed the importance of an authority figure who can provide psychological supervision.
In Sec~\ref{sec:visual}, we leveraged visually unobstructive virtual hands to represent the authority which were proven to be unrelated to performance.
In the future, we plan to further investigate how to effectively present an authority figure in the VR scene to enhance supervision.
This requires investigation into the visual representation of the authority (a complete therapist figure, an abstract icon, etc.) and how the authority is involved in the exercises (standing by, virtually pushing the patient's movements, etc.). 
\vspace{-3mm}
\section{Conclusion}
We investigated the current needs and challenges of self-monitored physical rehabilitation through clinical session observations and focus group interviews.
Based on this we presented design recommendations including providing motion awareness and guidance based on accurate tracking, asynchronous involvement of the therapists, providing multimodal supervision, automated progress tracking, and mobility.
We then designed a self-monitored physical rehabilitation system with VR and haptic feedback, \systemname, and further characterized the visual and haptic feedback features with two user studies. 
We found that the third pov and supplemental haptic motion guidance effectively enhances motion accuracy, awareness, and stability.
Finally, we implemented \systemname and evaluated its effectiveness in facilitating self-monitored physical rehabilitation through a clinical user study and semi-structured interviews.
The results showed that compared to screen displays without haptic guidance, \systemname could significantly improve motion stability in the short-term and was preferred by patients for at-home physical rehabilitation in the interviews.


\bibliographystyle{ACM-Reference-Format}
\bibliography{sample-base}

\end{document}